\documentclass{aa}
\usepackage{graphicx}

\begin{document}

\title
{A radio catalog of Galactic HII regions for applications from decimeter to 
millimeter wavelengths}

\author{ R.~Paladini\inst{1}, C.~Burigana\inst{2}, R.D.~Davies\inst{3}, 
         D.~Maino\inst{4,5}, M.~Bersanelli\inst{5,6}, B.~Cappellini\inst{5}, 
         P.~Platania\inst{5}, G.~Smoot\inst{7}}

\institute{
SISSA, International School for Advanced Studies,
via Beirut 2-4, I-34014 Trieste, Italy
\and
IASF, Istituto di Astrofisica Spaziale e Fisica Cosmica, 
Sezione di Bologna - Consiglio Nazionale delle Ricerche,
via Gobetti 101, I-40129 Bologna, Italy
\and
University of Manchester, Jodrell Bank Observatory,
Macclesfield - Cheshire SK11 9DL, UK
\and
Osservatorio Astronomico di Trieste,
via G. B. Tiepolo 11, I-34131 Trieste, Italy
\and
Universit\`a degli Studi di Milano, 
via Celoria 16, I-20133 Milano, Italy
\and
IASF, Istituto di Astrofisica Spaziale e Fisica Cosmica, 
Sezione di Milano - Consiglio Nazionale delle Ricerche,
via Bassini 15, I-20133 Milano, Italy
\and
LBNL, SSL, Physics Department, University of California,
Berkeley, CA 94720, USA
}

\offprints{R.~Paladini, paladini@sissa.it }
 
\date{Received July 27, 2002; accepted October 10, 2002}

\authorrunning{R.~Paladini et~al. }
\titlerunning{A catalog of Galactic HII regions}

\abstract{
\noindent
By collecting the information from
24 previously published lists and catalogs,
we produce a comprehensive catalog (Master Catalog)
of 1442 Galactic HII regions.
For each object, we quote the original fluxes and diameters
as well as the available information on
radio line velocities, line widths and line temperatures
and the errors on these quantitities.
References to the original works are also reported.
By exploiting all these data we produce
a Synthetic Catalog of fluxes and diameters
(with corresponding errors)
at 2.7~GHz. This choice is motivated by the
extensive, although not complete, information available
at this frequency,
widely spread among many different catalogs,
and by its relevance for both detailed studies
on Galactic HII regions and 
the extrapolation up to millimetric wavelengths.
The catalog can be used for detailed studies of Galactic
HII regions and, by extrapolation, for investigations of HII regions
up to millimetric wavelengths. In particular, 
we discuss the 
study of the effects of microwave emission from HII regions on the new
generation of Cosmic Microwave Background (CMB) experiments. 
We present
simulations of the detection of HII regions in the {\sc Planck} high
resolution CMB survey, and briefly analize some of the typical
applications of our catalog to the evaluation of CMB
anisotropy experiments such as calibration, beam reconstruction and
straylight effects. 
The Master Catalog and the Synthetic Catalog
are available via ftp at: cdsarc.u-strasbg.fr.
This work is related to {\sc Planck}-LFI activities.\\
\keywords{HII Regions  -- Catalogs -- cosmic microwave background}
}

\maketitle

\section{Introduction}
\noindent
HII regions are very important radio
sources within our Galaxy. They
not only provide information about the early stages of stellar formation
but represent also a unique tool to investigate
the Galactic spiral structure. Moreover, their radio emission,
not attenuated by dust extinction, is   
powerful enough to enable us to probe distant parts
of the Galactic plane.
Despite their astrophysical relevance, there is no currently available
comprehensive catalog of these objects.
A few optical catalogs
(Sharpless 1959; Mars\'alkov\'a 1974; Fich $\&$ Blitz 1982)
do exist but they are severely limited by the
dust obscuration effects mentioned above.
At radio wavelengths,
the most comprehensive catalog of diffuse brightness ne\-bu\-lae
is the $COBE$-HII Catalog at 2.7~GHz (Witebsky 1978a, 1978b).
It includes
909 objects and has provided a reliable tool for e\-sti\-ma\-ting the
contribution
of HII regions to Galactic free-free in the analysis of the
$COBE$ data.
More recently, a list of 760
objects with angular diameter less than $10'$
has been compiled by Kuchar $\&$ Clark 1997 from the Green Bank (Condon et al. 1989)
and Parkes-MIT-NRAO (Condon et al. 1993; Tasker et al. 1994) radio surveys at 4.85 GHz.\\
These catalogs can now be combined and updated
with other recently published data.
A strong motivation supporting this work is
its relevance to
cosmic microwave background (CMB) studies.
Since the successful results of the $COBE$ satellite (Smoot et al. 1992),
increasing efforts have been devoted to investigating
the cosmological info\-rma\-tion contained in
the CMB anisotropies.
Recent balloon experiments such as
BOOMERANG (de Bernardis et al. 2000) and MAXIMA-1
(Hanany et al. 2000) as well as ground interferometer
experiments as DASI (Leitch et al. 2002), CBI (Mason et al. 2001)
and VSA (Scott et al. 2002)
have achieved remarkable results but, at
the same time, have shown the necessity to improve our knowledge
of all the relevant astrophysical foregrounds
which may affect the recovery of the CMB angular power spectrum
and, ultimately, of the cosmological parameters.  
In pa\-rti\-cular, the free-free emission is the least known among
the Galactic foregrounds.
However, much of the
free-free emission at low Galactic latitudes is produced by in\-di\-vi\-dual
bright HII regions.
Because of their strong emission, they may represent
a potential source of contamination
for the new generation of nearly full-sky coverage CMB anisotropy
space missions at high sensitivity and resolution.
These sources are also good candidates for receiver and antenna
pattern calibration.\\
By collecting the information from
24 previously published lists and catalogs,
we present in this paper a comprehensive catalog (Master Catalog)
of 1442 Galactic HII regions.
For each object, we quote the original fluxes and diameters
as well as the available information on
radio recombination line velocities, line widths and line
temperatures
and the errors on all these quantitities.\\
By exploiting these data we produce
a Synthetic Catalog of fluxes and diameters
(with corresponding errors)
at 2.7~GHz, motivated by the
extensive although not complete information available
at this frequency,
widely spread among the different catalogs,
and by its relevance for both detailed studies
on Galactic HII regions and 
the extrapolation up to millimetric wavelengths.
We discuss also some possible applications 
of this Synthetic Catalog to CMB studies.
The paper is organized as follows. Sect.~2 presents the stages
of the Master Catalog compilation (e.g., the 
selection criteria for the reference papers and the removal 
of non-thermal sources) and briefly summarizes
the content of the final compilation,
while an accurate and full description 
of the catalog content is given in Appendix~I.
Sect.~3 describes the production of the Synthetic Catalog at 2.7~GHz.
Sect.~4 concerns the completeness and flux limit of the
catalog and the spatial di\-stri\-bu\-tion of the listed sources.
Perspectives of possible applications of this Synthetic
Catalog to CMB studies are presented in Sect.~5.
Finally, we draw our main conclusions in Sect.~6.

\section{The Master Catalog compilation}

In this section we discuss the data
processing which was applied to convert the information
from the various re\-fe\-ren\-ce catalogs into a uniform format for
easy inter\-com\-pa\-ri\-son in a Master Catalog.

\subsection{Selection of the original catalogs}

The Master Catalog was compiled from the 24 radio catalogs shown in Table~1.
We follow the classification of HII regions
by Lockman et al. 1996 and only include catalogs of diffuse and
compact HII regions. Surveys of ultra-compact (UCHII)
and extremely extended objects (EHE) will
be considered in a forthcoming extension of the present list:
they are bright in the infrared (UCHII) and in the optical (EHE) 
but relatively weak at radio frequencies.

\noindent
The reason for following this route is driven
by the general goals of the work. Planned
and on-going CMB experiments
typically have angular resolutions ranging from a few arcmin
up to tens of arcmin in order to exploit the wealth
of cosmological information encoded
in the CMB anisotropies.
Therefore, diffuse and compact HII regions
-- with a typical angular extension of a few arcmin -- are the
most important to be taken into account if we want to investigate
the free-free emission as a contaminant of the CMB.\\
We have therefore considered
single-dish medium-resolution observations, typically
with beamwidths of a few arcmin, rather than
interferometer or synthesis telescope surveys
(Becker et al. 1994; Taylor et al. 1996)
which usually achieve a resolution of a few arcsec.
Also, single-dish low-resolution observations
(Westerhout 1958; Wilson $\&$ Bolton 1960),
whose typical angular resolution is $\sim 1^\circ$,
have not been included in the sample of selected
references in order to avoid problems arising
from double-counting of the sources.
Likewise we have not considered surveys simply oriented to the study of the
morphology and spectral behaviour of the diffuse Galactic radiation at centimetric
wavelengths
(Reich et al. 1984;
Reich et al. 1990; F\"urst et al. 1990a; F\"urst et al 1990b;
Duncan et al. 1995).\\

\begin{table*}[ht]
\caption[InpVal]
{List of the references (in alphabetical order) for the catalog.  
Listed
$l_{min}$ and $l_{max}$ provide the longitude range spanned by each
survey. The
range in latitude is $|b| \leq 2^\circ-4^\circ$. \\
$\dag$ Number of
sources after
subtraction of nonthermal objects. Notes: (a) the list of sources is retrieved by
the intersection with
the radio recombination line survey by Lockman et al. 1989; (b) early Parkes
survey at 2.7 GHz; (c)
Effelberg 100-m survey at 2.7 GHz; (d) Cygnus X; (e) nonuniform sky coverage; (f)
Green Bank and
Parkes-MIT-NRAO surveys cover
almost the whole sky: HII regions have been     
identified from either optical or radio recombination line surveys along
the disk.}
\begin{center}
\begin{tabular}{lcccccc}
\hline
\hline
{\hskip 1.1truecm {\it Reference }}              &    {\it $l_{min}$ }
&       {\it $l_{max}$}     &    {\it $\nu$ (GHz) } &
HPBW ($'$)            &   $Number^\dag \hskip 0.1truecm of \hskip 0.1truecm sources$   \\
\hline
Altenhoff et al. 1970          &    $335^\circ$ &  $75^\circ$    &
1.4/2.7/5       &
  10            &          325          \\
Altenhoff et al. 1979$^{(a)}$   &    $2^\circ$ &  $60^\circ$    &
5             &
2.6             &          265           \\
Beard 1966$^{(b)}$             &    $331^\circ$ & $333^\circ$   &
2.7          &
   7.4           &           13            \\ 
Beard $\&$ Kerr 1969$^{(b)}$   &    $27^\circ$ & $38^\circ$    &
2.7          &
   7.4           &           34             \\
Beard et al. 1969$^{(b)}$      &   $345^\circ$ &  $5^\circ$    & 
2.7          &
   7.4           &           72             \\  
Berlin et al. 1985             &    $4^\circ$   & $10^\circ$      &  
3.9          &
0.8$\times$18.3    &           45           \\
Caswell $\&$ Haynes 1987            &   $190^\circ$   &  $40^\circ$   &
5            &
   4.2           &          308              \\ 
Day et al. 1969$^{(b)}$        &  $307^\circ$  &  $330^\circ$   &
2.7          & 
  8.2           &          109               \\ 
Day et al. 1970$^{(b)}$        &  $37^\circ$   &  $47^\circ$     &
2.7          &
   8.2           &           48               \\
Downes et al. 1980            &  $357^\circ$ &  $60^\circ$      &
5            &
   2.6           &          169              \\
Felli $\&$ Churchwell 1972     &    ${(e)}$      &   ${(e)}$          &
   1.4       &
   10            &           80             \\
F$\ddot{u}$rst et al. 1987$^{(c)}$    &    ${(e)}$       &   ${(e)}$          &
2.7          &
4.27             &            7            \\  
Goss $\&$ Day 1970$^{(b)}$     &    $6^\circ$ &   $26^\circ$    &
2.7          &
   8             &           85               \\
Kuchar $\&$ Clark 1997         &      ${(f)}$   &    ${(f)}$        &
4.8          &
  3.1/4.2          &          760 \\
Mezger $\&$ Henderson 1967     &      ${(e)}$   &    ${(e)}$          &
 5            &
     6.3           &           17             \\
Reich et al. 1986$^{(c)}$      &      ${(e)}$   &    ${(e)}$          &
2.7           &
4.27             &              5             \\
Reifenstein et al. 1970        & $348^\circ$  &  $80^\circ$     &
5            &
   6.5           &          105             \\
Thomas $\&$ Day 1969a$^{(b)}$  & $288^\circ$  &  $307^\circ$    &
2.7          &
   8.2           &           39            \\
Thomas $\&$ Day 1969b$^{(b)}$  & $334^\circ$  &  $345^\circ$    &
2.7          &
   8.2           &           29             \\
Wendker  1970$^{(d)}$          & $76^\circ$   &  $84^\circ$      &
2.7          &
  11            &           77               \\
Wilson et al. 1970             &  $282^\circ$ &  $346^\circ$   &
5            &
   4
       &          132            \\
Wink et al. 1982               &     ${(e)}$   &      ${(e)}$          &
 5/15/86       &
   2.6/1/1.3        &          112             \\
Wink et al. 1983               &  $359^\circ$  & $50^\circ$    &
14.7          &
    1          &           84            \\
\hline
\hline
\end{tabular}
\end{center}
\end{table*}

\subsection{Identification and removal of SNRs}

The selected catalogs may contain in principle 
also some Galactic sources, 
different from HII regions, that we need to exclude from
our Master Catalog compilation.
Therefore, supernova remnants (SNRs)
of each selected catalog have been identified 
by comparing the coordinates of the catalog sources with those of 
the catalog of Galactic supernova remnants by Green 2000. 
The identification has been performed 
taking into account the average position uncertainty 
of the selected catalogs of HII regions 
and of the Green Catalog (both $\sim 1'$).

\subsection{Quoted flux densities and angular diameters}

\noindent
We provide the flux densities integrated over each source, $S$,
and the source angular diameter, $\Theta_{HII}$. 
In a few re\-fe\-ren\-ces (Wink et al. 1982 at
86 GHz; Wink et al. 1983) the flux density
is given only as a peak value. 
We correct this to a true
value by assuming that 
the source has a Gaussian profile and using the observed
diameter (see Rohlfs 1990).\\
\noindent
The source angular diameter
is given in terms of observed diameter in a few references (Altenhoff et al.
1979, Downes et al 1980; Wink et al. 1983).
We derive a true diameter again assuming a Gaussian
source profile. For source dimensions   
significantly smaller than the beamwidth, the references
give upper limit diameters.
The Master Catalogs includes
these derived flux densities and diameters for
completeness.

{\subsection{Other relevant data}}

In addition to the basic data on flux density and diameter, for
each source there are available further relevant data which
make the catalog more useful.\\
We include the notes from
Kuchar $\&$ Clark 1997 on the environment of each source, indicating
whether the source is complex or if it has a strong source nearby.
Again, following Kuchar $\&$ Clark 1997 we give radio
counterparts in other catalogs other than those referred to in Sect.~2.1.
Optical counterparts of the unobscured sources are obtained
from the identification given in the Catalog by M\'arsalkov\'a 1974.\\
We include available radio recombination line
data because they provide important information on
kinematics, distances and electron density for each HII region.
The data are given by Downes et al. 1980, Caswell $\&$ Haynes 1987,
Kuchar $\&$ Clark 1997, Reifenstein et al. 1970, Wilson et al. 1970, 
Wink et al. 1982 and Wink et al. 1983 in the source
catalogs listed in Sect.~2.1 and are complemented by the
recombination line data of Lockman 1989.

\subsection{The Master Catalog}

All sources observed
at least by one of the surveys of Table~1 
and recognized as Galactic HII regions
were included in the
Master Catalog. 1442 sources are listed.
All relevant information from each of the reference catalogs is
retained. Data for individual HII regions are available
from the references at 8 frequencies (1.4, 2.7, 3.9, 4.8, 5., 14.7, 15
and 86~GHz). This large data base is presented in a readily accessible
form as 11 {\it sub-catalogs}:\\

\vspace*{0.1truecm}
\noindent
sub-catalog \hspace*{0.3truecm}  1     \hspace*{0.7truecm}  coordinates (epoch 2000) and remarks\\
sub-catalog \hspace*{0.3truecm} 2 / 3   \hspace*{0.4truecm}   flux density (Jy) / 1--$\sigma$ ($\%$) error \\
sub-catalog \hspace*{0.3truecm}  4 / 5   \hspace*{0.4truecm}  diameter (arcmin) / 1--$\sigma$ ($\%$) error \\
sub-catalog \hspace*{0.3truecm}  6 / 7   \hspace*{0.4truecm}  line velocity (km/sec) / 1--$\sigma$ error (km/sec)\\
sub-catalog \hspace*{0.3truecm}  8 / 9   \hspace*{0.4truecm}  line width (km/sec) / 1--$\sigma$ error (km/sec)\\
sub-catalog \hspace*{0.2truecm} 10 / 11  \hspace*{0.2truecm}  line temperature (K) / 1--$\sigma$ error (K).\\

\vspace*{0.2truecm}

\noindent
The sub-catalogs give, for each source, an identification
number along with the position in Galactic and celestial coordinates.
Apart from sub-catalog 1, each sub-catalog has 37 columns. The
columns are in order of increasing frequency of observation; at
each frequency the columns are in alphabetical order of the references
in Table~1. We note that for sub-catalog 6-11 (the radio
recombination line data) the observations may refer to frequencies other than
those in the main continuum catalog
(in this case, the line frequency is reported).\\
Appendix~I gives a detailed
description of the content of each sub-catalog.

\section{The Synthetic Catalog at 2.7~GHz}

The final task in bringing together the wide-spread radio
data on HII regions is to construct a readily accessible
catalog summarizing the basic information on each of the
1442 sources covered in the comprehensive data base of the
Master Catalog. This Synthetic Catalog gives the best
available data on flux density, angular diameter and, where
available, the line velocity.\\
2.7 GHz was chosen as the base frequency for the Synthetic
Catalog; it lies in the middle of the frequency range of
those catalogs containing most of the data, namely 1.4, 2.7 and 5 GHz.
Since there is not complete source co\-ve\-ra\-ge
at any one frequency, we derive the flux density at 2.7 GHz from the
observed frequencies for all the sources. We
now describe how best-estimate values of the flux density,
diameter and velocity with corresponding error
estimates are derived from the Master Catalog.

{\subsection{Flux density estimates at 2.7~GHz}}

\vspace*{0.2truecm}
\noindent
We estimate the flux density at 2.7~GHz in four different ways:\\
\begin{enumerate}
\item{we use the flux density measures at 2.7~GHz when these
are available;}
\item{if there are no flux density measures at 2.7~GHz and there
are available measures at only one other frequency, we extrapolate
these values to 2.7~GHz with the theoretical spectral index $\alpha =-0.1$
($S \propto \nu^{\alpha}$) which is typical of thermal bremsstrahlung 
emission in a thin plasma;}
\item{if there are flux density measures only at frequencies
$\ge$~14.7~GHz, we first select the lowest among the given
frequencies -- averaging over 14.7 and 15~GHz
when both are available -- and then extrapolate to 2.7~GHz with $\alpha =-0.1$;}
\item{if there are only  flux density measures at frequencies
other than 2.7 GHz and at least one measure at $\nu \le 5$~GHz, 
we follow these lines: we exclude the va\-lues at 3.9, 14.7,
15 and 86~GHz, we take the weighted average between 4.8 and 5 GHz
if both measures are available, we interpolate the
remaining values to 2.7 GHz (or extrapolate with $\alpha =-0.1$ 
if only one frequency measure is left).}
\end{enumerate}

\noindent
We consider the data at 3.9, 14.7, 15 and 86 GHz less
re\-lia\-ble for the following reasons: the 3.9 GHz measure 
may be significantly affected by the 
ellipticity of the antenna pattern; at 14.7 and 15 GHz the 
observing resolution is much higher than the typical reference
catalogs resolution; at 86 GHz, in addition to the higher
resolution, dust emission may contribute largely to the
observed flux. Moreover, we prefer to interpolate between 1.4 and 5 GHz
rather than extrapolate to 2.7 GHz directly from only one
of these two frequencies because we have no information
on the frequency location of the free-free knee.\\
In cases when multiple observations are available at the
same frequency, a weighted
mean flux density is computed, using the errors discussed below.

\begin{figure*}[ht]
\mbox{}
\caption{Results of the flux error estimate for: earlier Parkes survey at 2.7 GHz
(top left); Mezger $\&$ Henderson 1967 (top right); Wink et al. 1983 (bottom
center). Plotted points in each panel correspond to the $\sigma (S_{*})$ distribution (where
$\sigma$ is defined as $100(S_{*} - S)/S$).
Overlaid -- solid line -- is the best fit function.
See Sect.~3.1 for more details on the quantities.}
\centerline{\includegraphics[width=7cm,height=6cm]{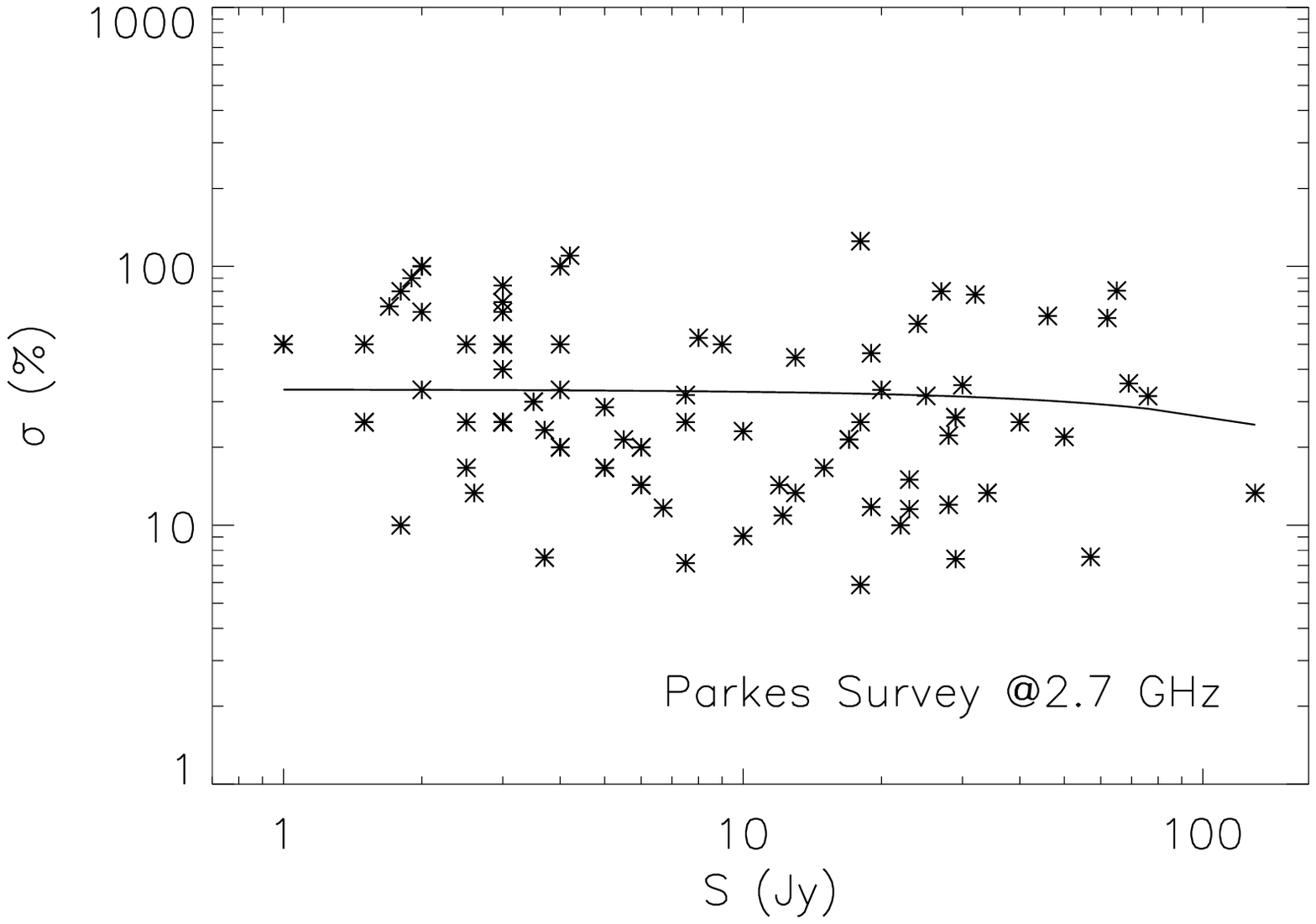}
            \includegraphics[width=7cm,height=6cm]{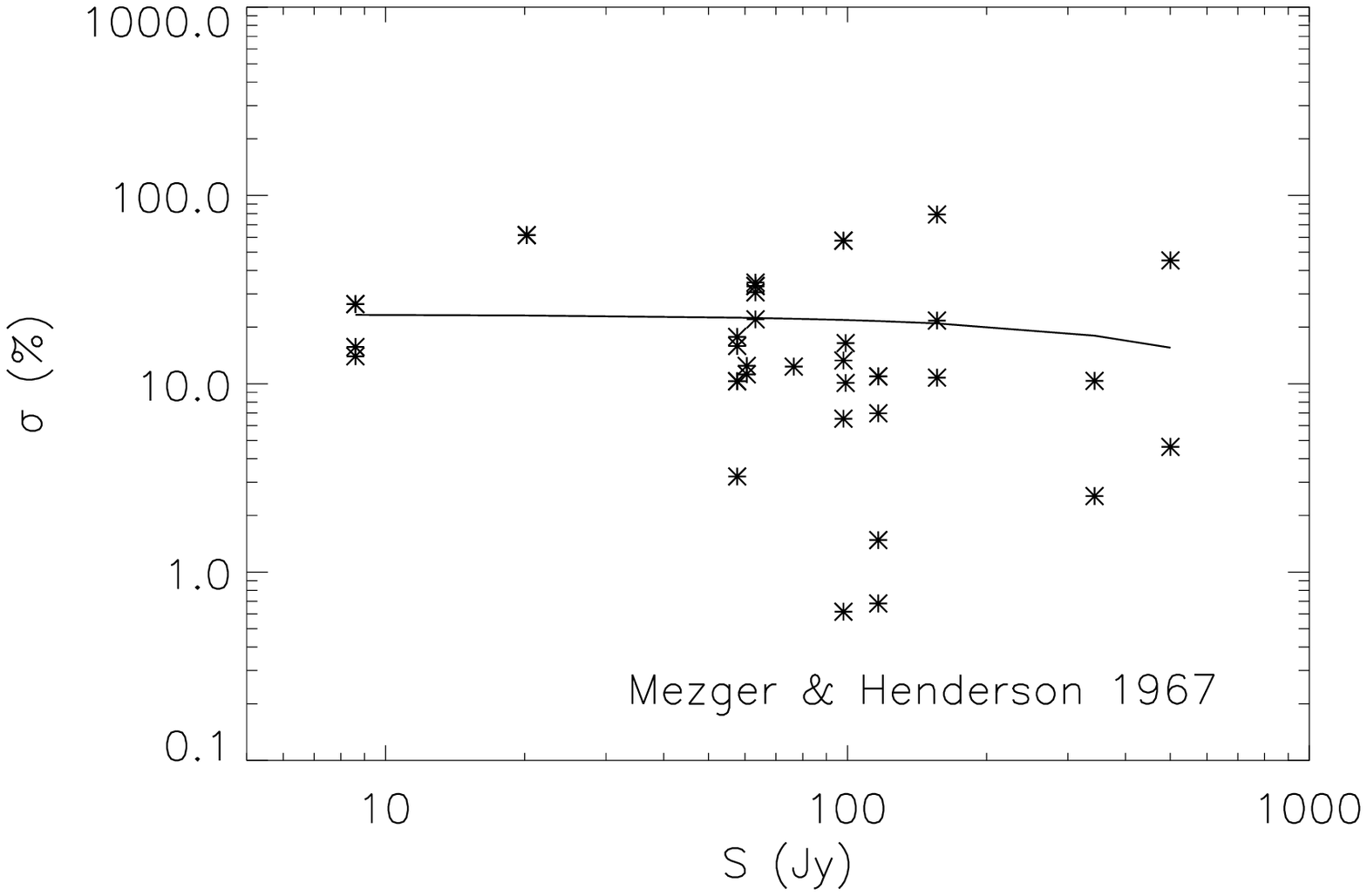}}
\centerline{\includegraphics[width=7cm,height=6cm]{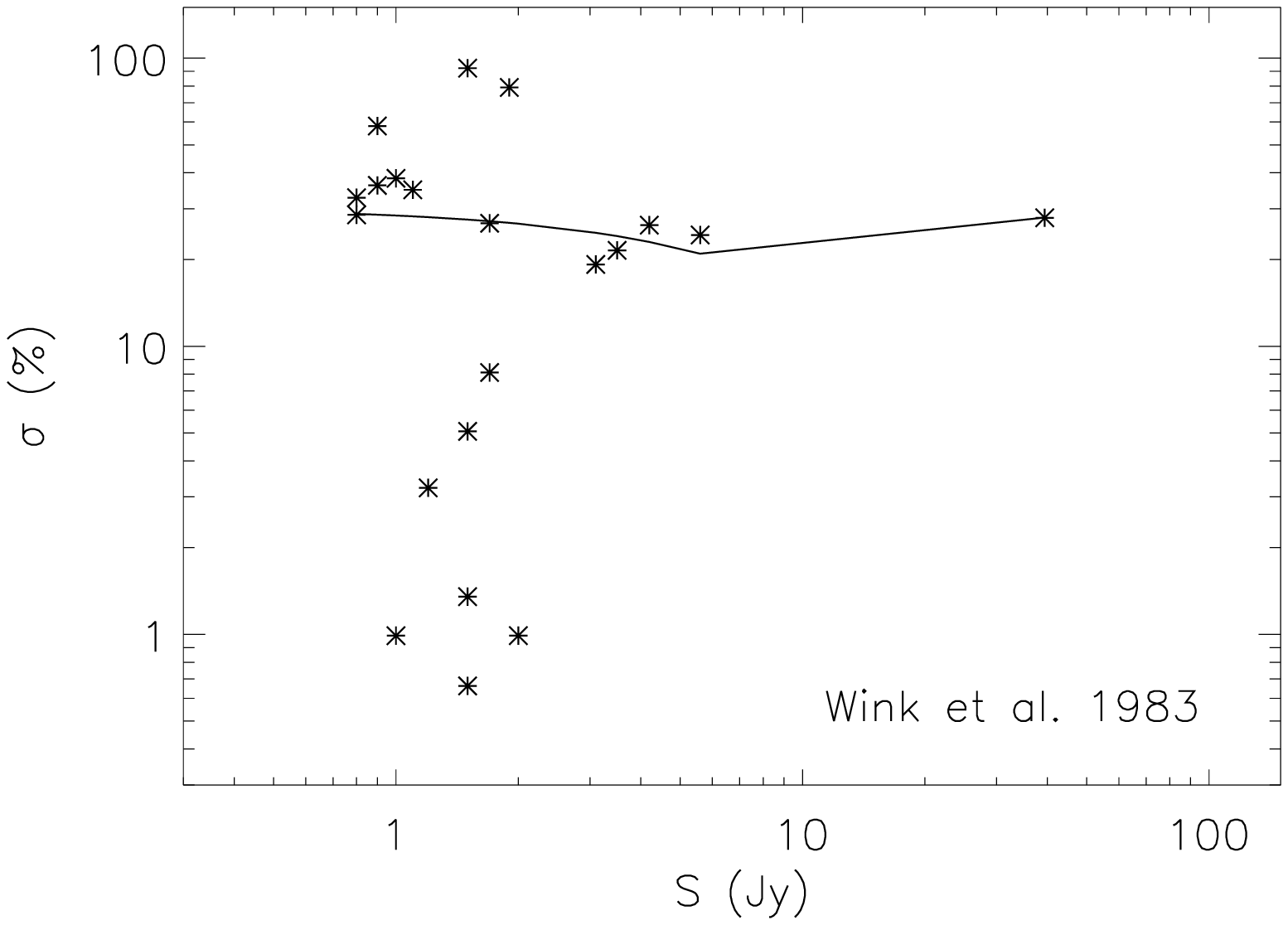}}
\end{figure*}

\noindent
For the majority of the
sources in the reference
catalogs, flux density errors are quoted and these are used in
de\-ri\-ving the flux density and its error in the Synthetic Catalog.
In the catalogs where no errors are given, an estimate of the error
is made by comparing the flux $S_{*}$ from a catalog without errors with
the flux $S$ from a catalog (or some catalogs) giving errors.
In particular, we compute the re\-la\-ti\-ve (\%) dispersion,
$\sigma = 100 (S_{*} - S)/S$,
and try to fit the resulting distribution
in the $S_{\*} - \sigma$ plane with
a con\-stant, $\sigma = a$ , a linear, $\sigma = a + b S_{*}$,
and a quadratic, $\sigma = a + b S_{*} + c S_{*}^{2}$,
dependence of $\sigma$ on $S_{*}$.
In these equations the fluxes are in Jy.
Before fitting the $\sigma (S_{*})$
distribution, we remove the relatively small number of points
with a dispersion $\simeq 130 \%$.
Although we have considered also fits with
a constant, linear, and quadratic
dependence of $\rm{log} \sigma$ on $\rm{log} S_{*}$,
we prefer to use
the distributions re\-co\-ve\-red from the fit carried out
with linear variables, since they
give, as expected, typical values of $\sigma (S_{*})$ larger than
those obtained in the case of logarithmic variables and
so giving more conservative error estimates. Fig.~1 shows the results
of the procedure we have described.\\
An example of the estimates of errors is given by the early Parkes
2.7 GHz catalogs. We compare the list of Parkes sources
to the reference catalogs at the same frequency -- 2.7~GHz -- 
(namely, Altenhoff et al. 1970,
F$\ddot{u}$rst et al. 1987, Reich et al. 1986, Wendker 1970),
in order to retrieve the subset of Parkes sources whose flux
has been quoted with an estimated error. In particular, the comparison 
retrieves 105 sources in common with the Altenhoff et al. 1970 ca\-ta\-log.
The scatter $\sigma$ in the flux density differences,  quite well fitted by the law
$\sigma \simeq 33.5-0.1\times S$, is comparable to the errors of 10--30~\% of the 
Altenhoff et al. catalog alone.\\
Similar comparisons have been carried out for the Mezger $\&$ 
Henderson 1967 data and for Wink et al. 1983 data. 
However, as for the Mezger $\&$ Henderson data, the comparison
with other single catalogs does not provide a si\-gni\-fi\-cant
statistical subset of sources. Therefore, we estimate the error
by considering simultaneously the sources from the Mezger $\&$ Henderson
catalog and from all the other catalogs at 5 GHz (which leads to
consider altogether 38 common sources).
For the Wink et al. 1983 catalog, we derive an error estimate 
from the comparison with the Wink et al. 1982 catalog for
which we have 21 common sources.\\ 
After having estimated the errors on the flux density measures for all catalogs, 
we compute the errors on the flux densities of the Synthetic Catalog at 2.7~GHz. 
When an extrapolation with 
a spectral index $\alpha =-0.1$ from a measure at a single frequency different
from 2.7~GHz is applied, we use the standard error propagation rules. 
No error on $\alpha$ is included in the error propagation,
for simplicity. \\
Table~3 summarizes the errors quoted in the original ca\-ta\-logs and those 
estimated in this work.

\subsection{Angular diameter}

We discuss here the assignment of an angular diameter with its
associated error for every source in the Synthetic Catalog. The
published diameter data are not as comprehensive as the flux
density data described in Sect.~3.1; nevertheless we will assign
a value of the diameter and an error for each source. For 42~\% 
of the sources (derived from the Parkes catalogs and those of
Mezger $\&$ Henderson 1967, Wendker 1970, Wink et al. 1982, 
Wink et al. 1983, Berlin et al. 1985) there are no listed
diameter errors. For 14 ~\% of the sources no diameter is
given; for these we derive an indicative diameter.

\noindent
We assign a diameter  for sources in the Synthetic Catalog
as follows:

\begin{enumerate}
\item{when 2.7 GHz diameters are given, we use these values (possibly 
their weighted average);} 
\item{if there is at least one diameter measure at 5~GHz or at 4.8 GHz,
we neglect possible measurements at other frequencies
and use this value or, alternatively, the weighted average of the
diameter measures at 4.8 and 5 GHz when available,
using the errors di\-scus\-sed below;}
\item{if diameter data are given only at frequencies 
other than $\sim 4.8-5$~GHz,
we use the weighted average of the available frequencies, after
previous exclusion of the measures at $\nu \sim 14.7$~GHz,
using the errors discussed below;}
\item{if no diameter data is given  
at frequencies lower than $\sim 14.7$~GHz, 
we use the available measure at 14.7~GHz or at 15~GHz
or, alternatively, their weighted average when both values are
provided using the errors discussed below
(we remember that no diameter data are available at 86~GHz).}
\end{enumerate}

\noindent
Values at 14.7 and 15 GHz are preferably excluded because of the
much higher resolution at these frequencies than the 
typical reference catalog resolution.
In each case, when multiple
observations are available at the same frequency, a weighted mean
is computed, using the errors discussed below.\\
Finally, we consider the assignment of a diameter to the
14~\% of HII regions in the Synthetic Catalog which have
no quoted diameter in the reference catalogs. 
It is possible to derive a first-order indicative 
diameter by noting that the flux density and the diameter of HII regions are
weakly correlated, as shown in Fig~2 which includes
all the sources at 2.7~GHz with measured diameters. 
The best fit to the data gives 
as indicative diameter $\theta = 2.25 \times S$ with a typical error 
of a factor of 2--3. Each source with an indicative diameter is
annotated in the Synthetic Catalog; such diameter data clearly should not
be used in astrophysical analyses of the catalog.

\begin{figure}[h]
\mbox{}
\centerline{\includegraphics[width=7.5cm,height=5.5cm]{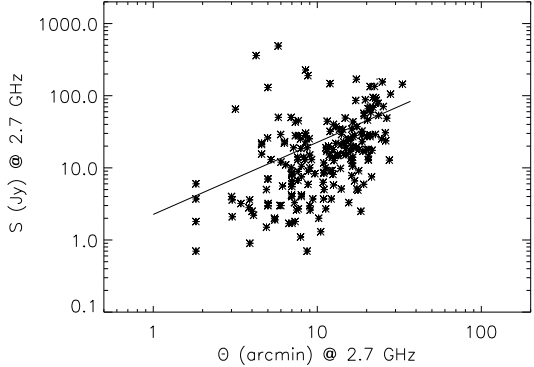}}
\caption[]{Correlation between flux densities at 2.7 GHz and angular
diameters observed at the same frequency. The solid line represents the
best-fit for the plotted distribution.}
\end{figure}
\vspace*{-0.5truecm}
\noindent
Following the same strategy as in Sect.~3.1, 
where a catalog has no quoted diameter errors we estimate
an overall error for that catalog by comparing
the observed diameters of that catalog with another
catalog having an adequate number of sources in common. 
However, with respect to the flux density case, we can now 
relax the strict requirement of comparing only different datasets 
at the same frequency. 
As shown in Fig.~3, there is no significant frequency dependence in the
measured diameters at 2.7~GHz (the Parkes survey)
and at 5~GHz (Altenhoff et al. 1970, Mezger $\&$ Henderson 1967 and
Reifenstein et al. 1970).
On the other hand, provided that an adequate number of sources in common
is retrieved, we prefer to consider in the comparison 
surveys at the same frequency and with similar
angular resolution. 
Therefore, let $\Theta_{*}$ ($\Theta$) be the source diameter measure 
in the catalog without quoted errors (with quoted errors).
We compute the relative (\%) dispersion $\sigma$
of the diameter measures, 
$\sigma = 100 (\Theta_{*} - \Theta)/\Theta$,
and try to fit the resulting distribution
in the $\Theta_{*} - \sigma$ plane 
again with a constant, linear, and  quadratic, 
dependence of $\sigma$ on $\Theta_{*}$ 
(or of ${\rm log} \sigma$  on ${\rm log} \Theta_{*}$;
in this case the fit error estimates are again less conservative) and after
removing the points with a dispersion $\simeq 130\%$.
In these equations 
the angular diameters are in arcmin.  

\begin{figure}[h]
\mbox{}
\centerline{\includegraphics[width=7.5cm,height=5.5cm]{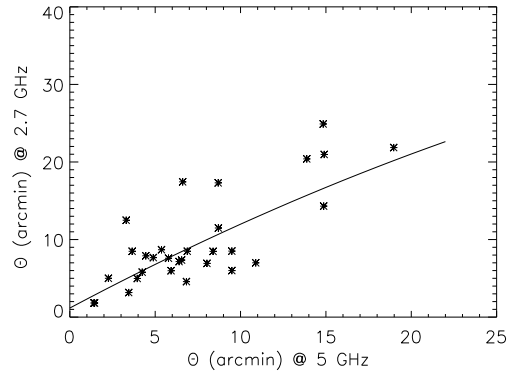}}
\caption[]{Diameter measures at 2.7~GHz (y-axis) versus
diameter measures and at 5~GHz (x-axis). 
It is evident the good agreement between the
two datasets. The best fit function is:
$\Theta_{\rm 2.7 GHz} = 1.17+1.17 \times \Theta_{\rm 5 GHz}$.
Data are from: earlier Parkes survey (2.7 GHz data); Altenhoff et al. 1970$^{a}$, 
Mezger $\&$ Henderson 1967; Reifenstein et al. 1970 (5 GHz data).}
\end{figure}

\noindent
Fig~4 shows the results of the error estimates. 
In pa\-rti\-cu\-lar, we consider the comparison between the following
datasets (the first one is that for which we are e\-sti\-ma\-ting the errors):
Altenhoff et al. 1970$^{a}$ vs. Reifenstein et al. 1970 (48 common 
sources); 
Altenhoff et al. 1970$^{b}$ vs. Kuchar $\&$ Clark 1997 (63 common sources); 
Caswell $\&$ Haynes 1987 vs. Wilson et al. 1970 (79 common sources); 
early Parkes 2.7 GHz Survey vs. Reifenstein et al. 1970 (31 common sources); 
Felli  $\&$ Churchwell 1972 vs. Kuchar $\&$ Clark 1997 (21 common sources);
Wendker 1970 vs. Reifenstein et al. 1970 (8 common sources); 
Wink et al. 1983 vs. Downes 1980 (59 common sources); Wink et 
al. 1982 vs. Altenhoff et al. 1979 (53 common sources).\\
As for the Altenhoff et al. 1970 catalog, the data have been
split because the diameters have been
measured in two different ways: either they were taken from a survey
of Galactic sources made at 5 GHz with the 140-ft 
antenna - beamwidth $6'$ 
(W.J.~Altenhoff, P.G.~Mezger and J.~Schraml, private communication) -
or they were measured directly from contour maps.
We will annotate the set of sources for which
the size was measured in the 5 GHz survey Altenhoff et al. 1970$^{a}$
and the remaining sources Altenhoff et al. 1970$^{b}$.
For the Berlin et al. 1985 catalog, since none of the comparisons 
with other catalogs retrieves
a significant number of common sources, we derive the error from the
mean of the dispersions $\sigma$ given by each comparison. 
Table~3 summarizes the estimated and quoted errors used 
in calculating the diameters in the Synthetic Catalog.

\begin{figure*}[p]
\centerline{\includegraphics[width=7.3cm,height=6.2cm]{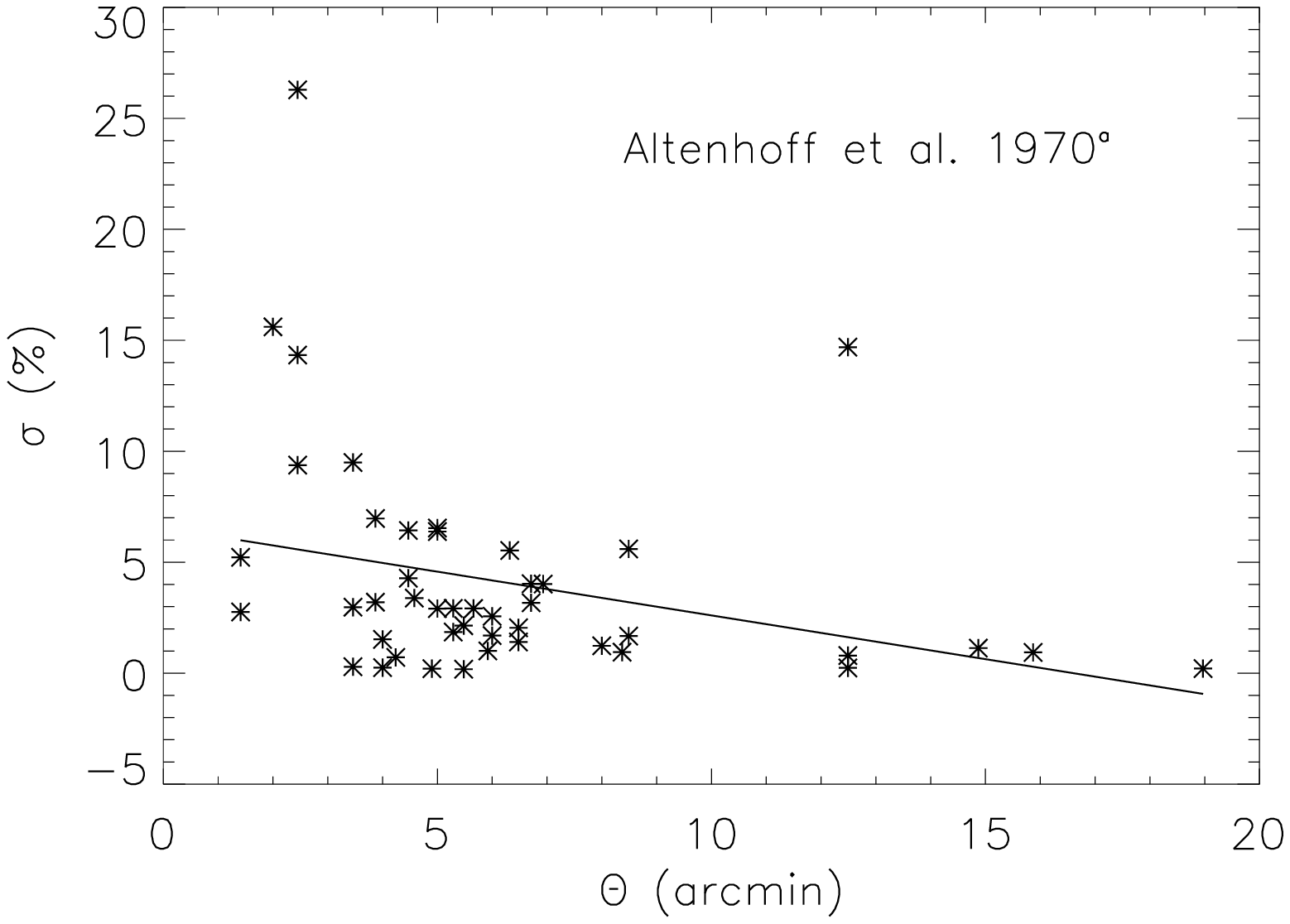}
            \includegraphics[width=7.3cm,height=6.2cm]{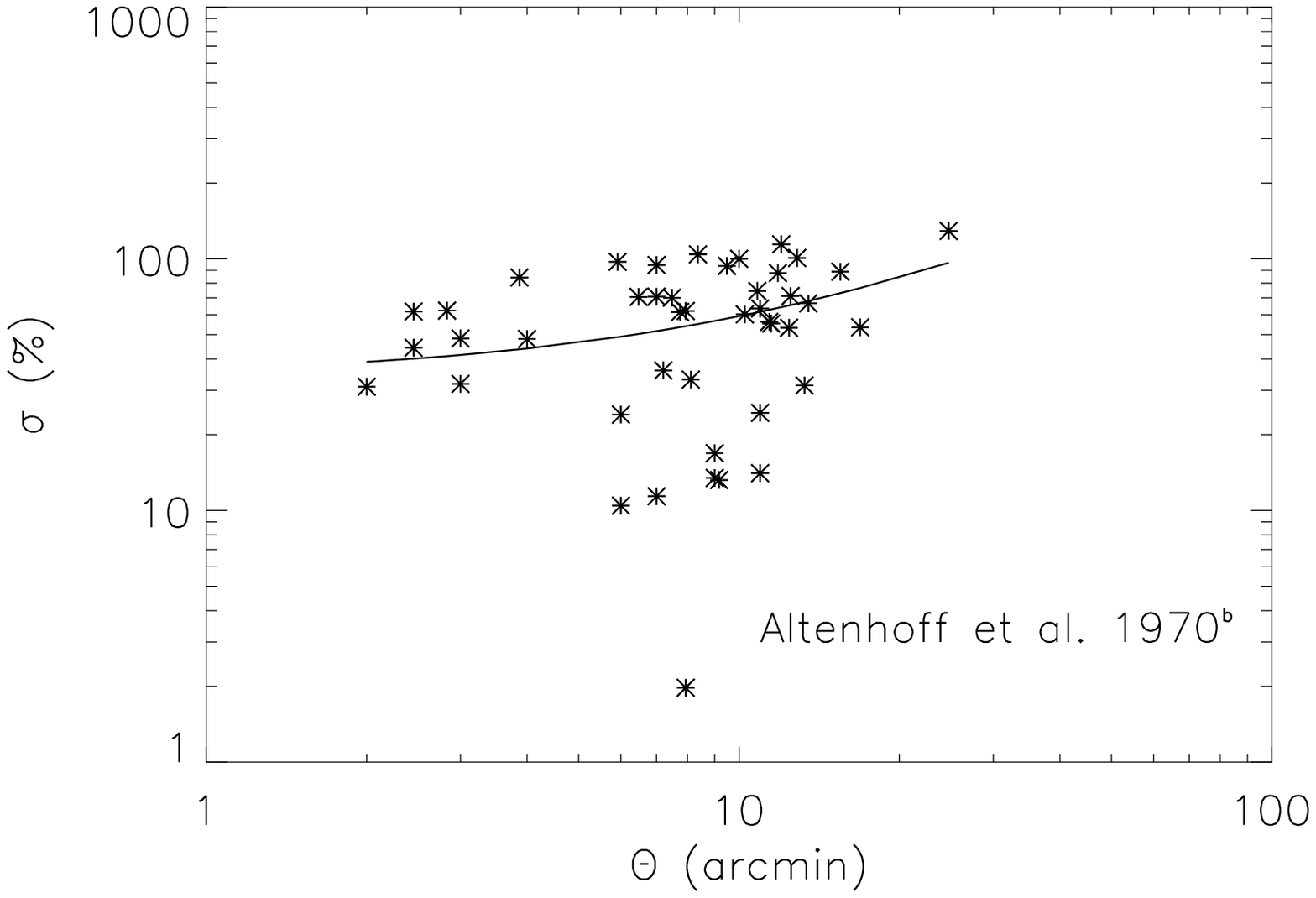}}
\centerline{\includegraphics[width=7.3cm,height=6.2cm]{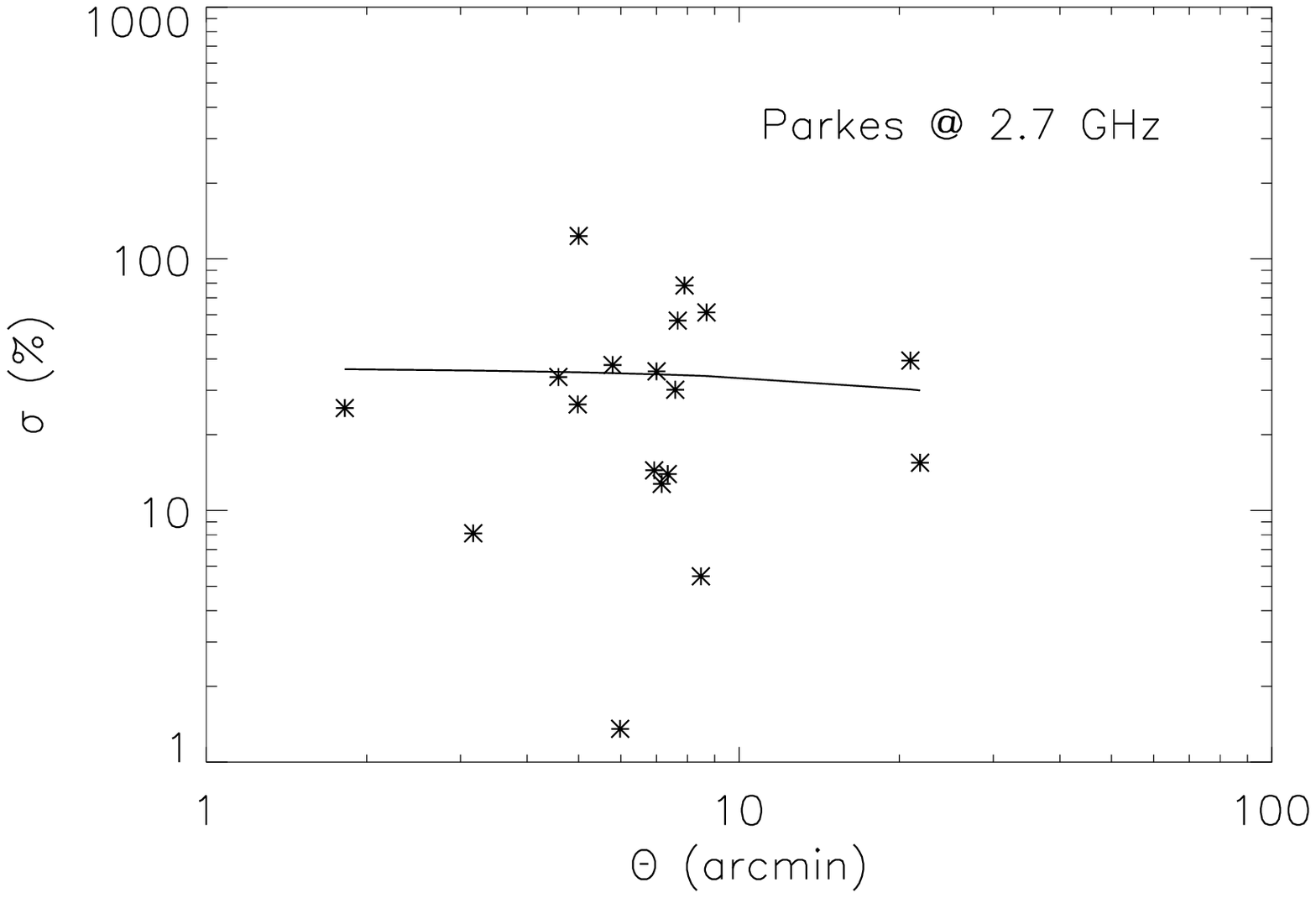}
            \includegraphics[width=7.3cm,height=6.2cm]{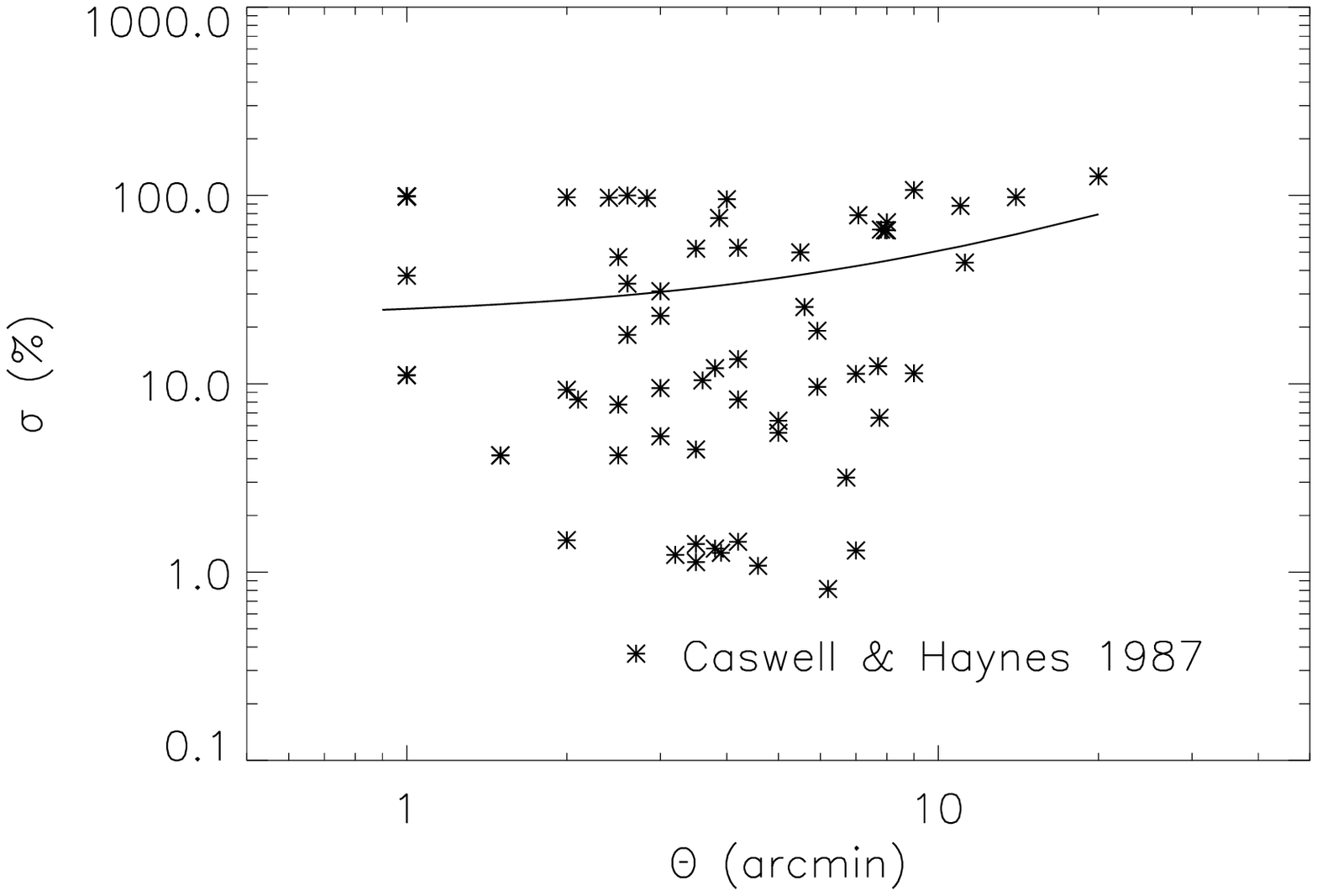}}
\centerline{\includegraphics[width=7.3cm,height=6.2cm]{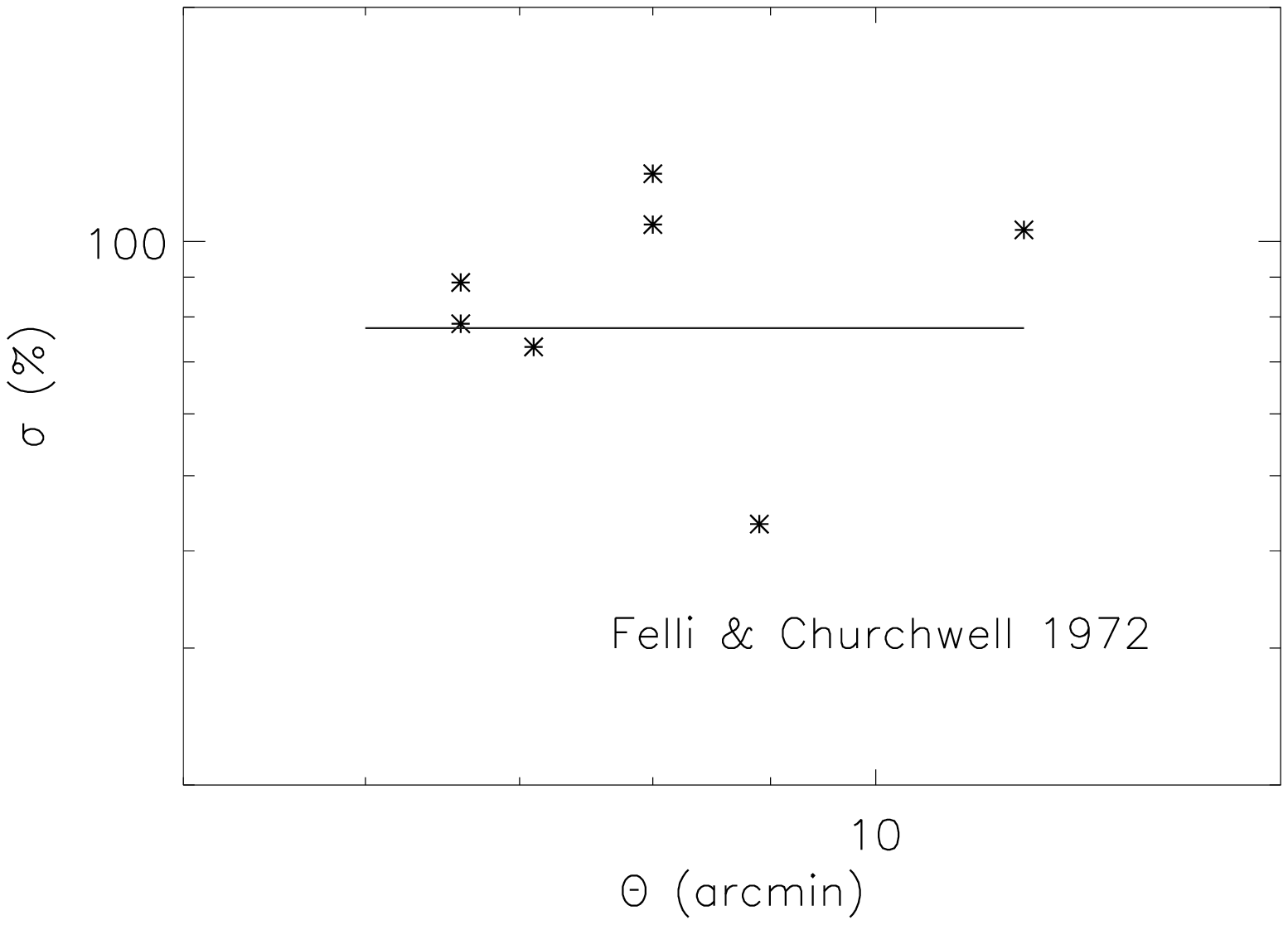}
           \includegraphics[width=7.3cm,height=6.2cm]{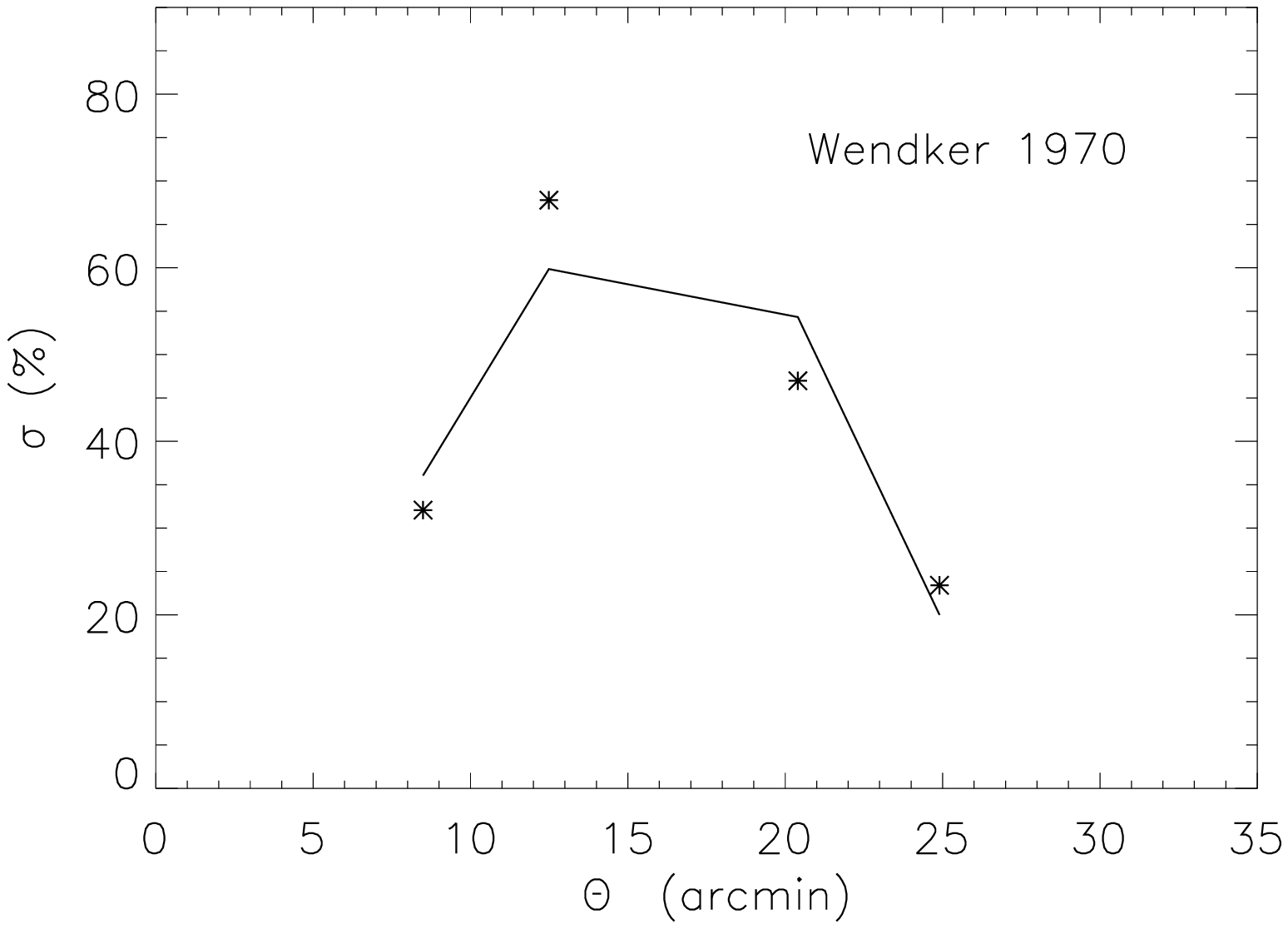}}
\centerline{\includegraphics[width=7.3cm,height=6.2cm]{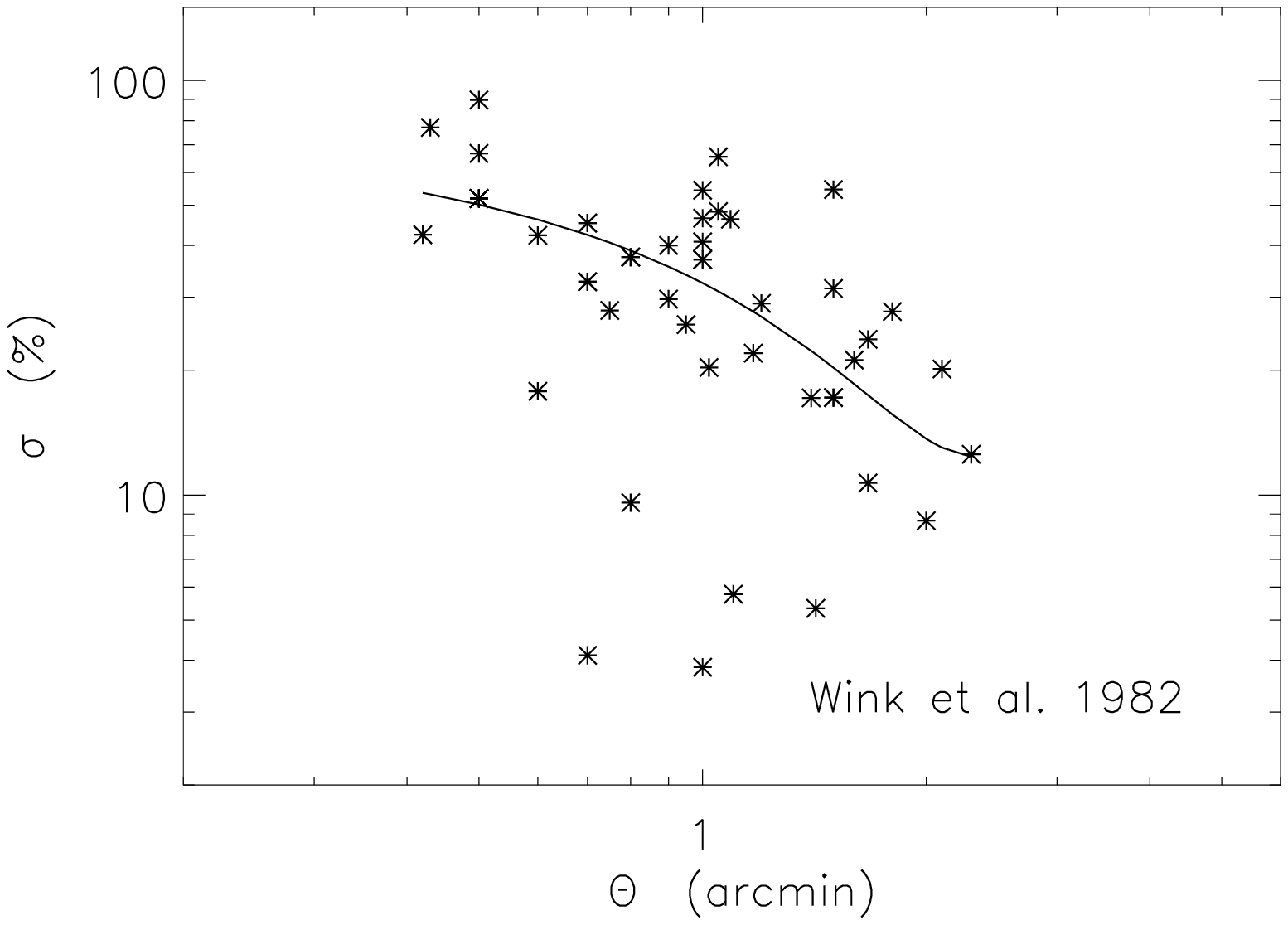}
           \includegraphics[width=7.3cm,height=6.2cm]{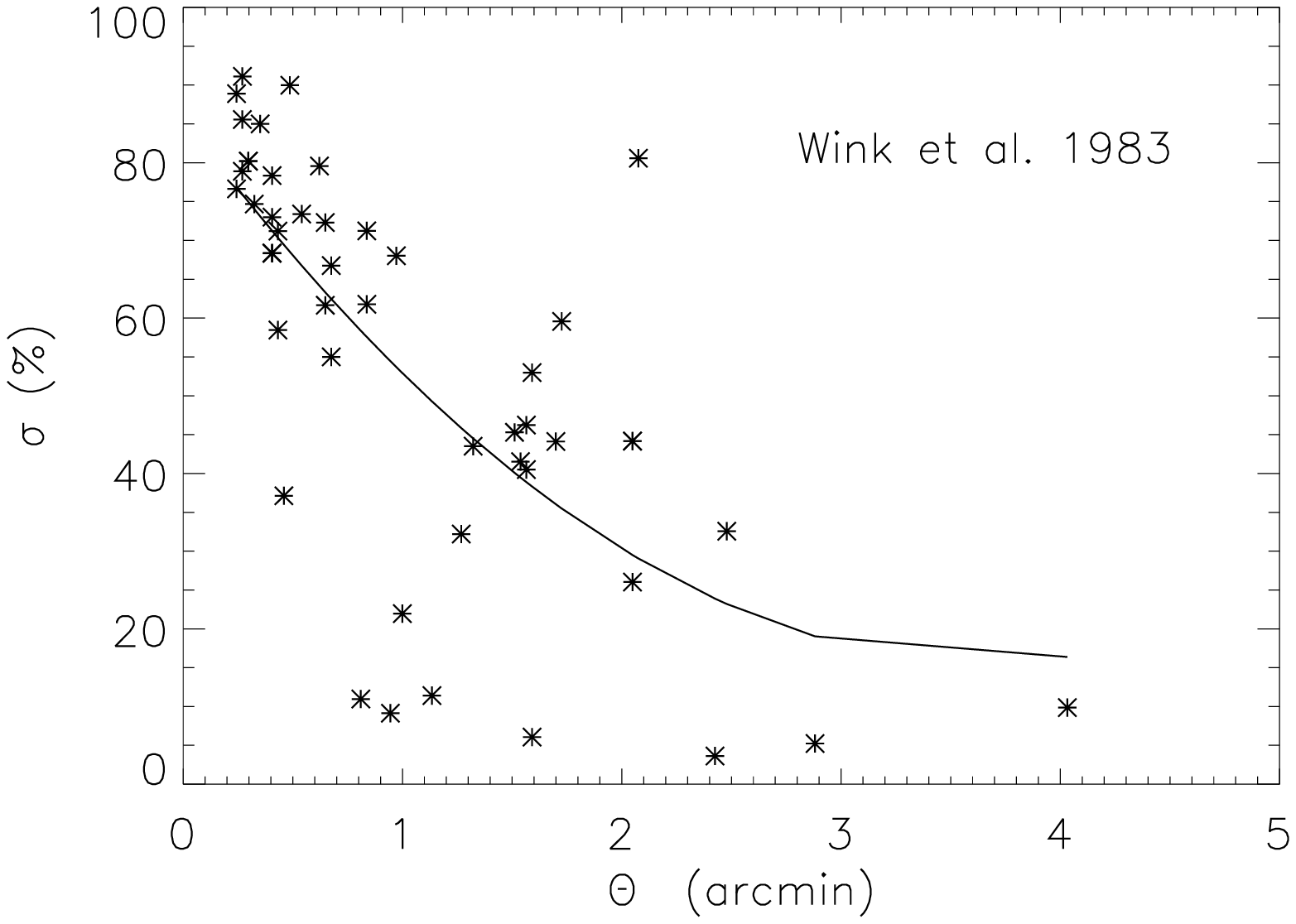}}
\caption[]{Results of the diameter error estimate for: Altenhoff et al.
1970$^{a}$
(top left); Altenhoff et al. 1970$^{b}$ (top right); earlier Parkes survey
at 2.7 GHz
(left, second row); Caswell $\&$ Haynes 1987 (right, second row); Felli
$\&$ Churchwell 1972
(left, third row); Wendker 1970 (right, third row); Wink et al. 1982
(bottom left); Wink et al.
1983 (bottom left). Plotted points in each
panel correspond to the $\sigma (\Theta_{*})$ distribution (where
$\sigma$ is defined as $100(\Theta_{*} - \Theta)/\Theta$).
Overlaid -- solid line -- is the best fit function.
See Sect.~3.2 for more details on the quantities.}
\end{figure*}

\subsection{Content of the Synthetic Catalog}

The Synthetic Catalog summarizes the known radio  
frequency information on 1442 Galactic HII regions. It contains
the position, flux density, diameter data for each HII region,
supplemented by velocity data where available. 
To those HII region with no published diameter data, an indicative diameter
is given (marked by $**$) on the basis
of the flux-size correlation in Fig~2. For sake of clarity, 
the first ten lines of the Synthetic Catalog are reported in Table~2.
The line velocity value that we quote in Col. 9 is
the weighted average of the available data (see Sect.~2.4 for details).
Although the original measures can be, for the same source, at different
frequencies, the wei\-gh\-ted a\-ve\-ra\-ge of these data is a meaningful
quantity and provides a useful first-sight kinematic
indication. Since the line velocity is an effect of  
the Galactic rotation motion, it does not strongly depend
on the frequency of observation.

\begin{table*}[ht]
\caption[InpVal]
{A selection of sources from the Synthetic Catalog is shown.
The arrengement of column is as follows:\\
\noindent
- Col. 1: source-numbering (records from 1 to 1442)\\
- Col. 2-3: galactic coordinates, l and b\\
- Col. 3-4: celestial coordinates (epoch 2000)\\
- Col. 5-6: derived 2.7~GHz flux density and 1--$\sigma$ error (Jy)\\
- Col. 7-8: derived angular diameter and 1--$\sigma$ error (arcmin)\\
- Col. 9-10: velocity relative to the LSR and 1--$\sigma$ error (km/sec)\\
- Col. 11: notes on individual sources (see Sect.~2.4 and Appendix I for details)\\
}
\begin{tabular}{ccccccccccccc}
\hline
\hline
N               &    {\it $l$ }
&       {\it $b$}     &    RA      &
DEC     &   S &   $\sigma_{S}$
&       $\theta$ & $\sigma_{\theta}$ &   $V_{LSR}$
&       $\sigma_{V_{LSR}}$ &  Notes    
\\ 
\hline
   1  &  0.1  &  0.0  &  17 45 51.3  &  -28 51  08  &   230.0  &  24.1  &    5.9  &     0.5  &  -27.4  &  4.0  &  W24 \\
   2  &  0.2  & -0.1  &  17 46 29.0  &  -28 49  07  &  209.4   &  10.5  &   10.7  &     0.5  &   24.5  &  3.5  &          \\  
   3  &  0.2  & -0.0  &  17 46 05.6  &  -28 46  00  &  177.6   &  38.1  &    9.2  &     0.5  &  -12.7  &  3.5  &      W24 \\
   4  &  0.3  & -0.5  &  17 48 17.0  &  -28 56  25  &    2.5   &   0.7  &    2.7  &     1.7  &   20.0  &  4.9  &         C,S \\
   5  &  0.4  & -0.8  &  17 49 41.7  &  -29 00  33  &    8.0   &   2.6  &    7.0  &     2.3  &   20.0  &  4.9  &         C,S \\
   6  &  0.4  & -0.5  &  17 48 31.1  &  -28 51  17  &    4.1   &   1.0  &    3.9  &     1.8  &   24.0  &  4.9  &         C,S \\
   7  &  0.5  & -0.7  &  17 49 32.2  &  -28 52  19  &    2.9   &   0.9  &    2.3  &     1.4  &   17.5  &  4.9  &           C \\
   8  &  0.5  & -0.1  &  17 47 11.6  &  -28 33  44  &   28.3   &   4.0  &    3.3  &     1.4  &   45.8  &  5.0  &          S,X \\
   9  &  0.5  &  0.0  &  17 46 48.2  &  -28 30  37  &   40.3   &   8.6  &    4.8  &     0.3  &   47.1  &  2.0  &              \\
  10  &  0.6  & -0.9  &  17 50 33.3  &  -28 53  19  &    2.5   &   0.8  &    2.4  &     1.3  &   15.0  &  4.9  &    S,S21,RCW142 \\
\hline
\hline
\end{tabular}
\end{table*}

\section{Statistical properties of the Synthetic Catalog}

\noindent
The Synthetic Catalog combines the observational data from 24 catalogs
with various flux density limits and different angular resolutions. 
The Kuchar $\&$ Clark 1997 ca\-ta\-log, for example, reaches to a sensitivity of
30 mJy but is not sensitive to sources with diameters greater than $\sim$10$^{\prime}$
due to its beamswitching strategy. 

\begin{figure}[h]
\mbox{}
\caption[]{Cumulative counts, N($> S$), for the Synthetic Catalog
(solid line) and 5 contributing catalogs: Kuchar $\&$ Clark 1997 (dash-dot line),
early Parkes 2.7 GHz catalogs (dashed line), Caswell $\&$ Haynes 1987  
(long-dash line), Altenhoff et al. 1970 (dotted line), Altenhoff et~al.
1979 (three dots-dashes).}
\centerline{\includegraphics[width=7cm,height=5.5cm]{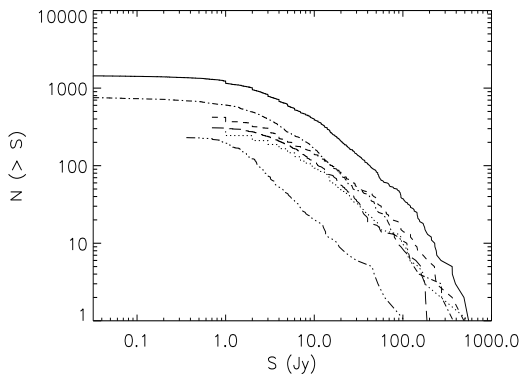}}
\end{figure}

\noindent
Other surveys (e.g. Altenhoff et al., 1970)
are sensitive to a wide range of angular scales but include only sources
stronger than 1 Jy. We now consider the properties of the sources
listed in the Synthetic Catalog in the light of these
selection criteria.

\subsection{Completeness and flux density limit}

\noindent
A useful way of investigating the completeness of a catalog
is to plot the integral count N($> S$) as a function of flux density $S$.
Fig~5 shows the integral count for the 1442 sources included
in the Synthetic Catalog and for the five contributing catalogs 
containing more than 200 HII regions. 
It is evident that the
Synthetic Catalog is losing sources fainter than $\sim$~1~Jy. The
data in the Synthetic Catalog for $S<1$~Jy are mainly contributed
by the Kuchar $\&$ Clark 1997 catalog. Some 50~\% of the
sources at 1~Jy are contributed
by this catalog, with a higher fraction at lower flux densities.

\begin{figure}[ht]
\mbox{}
\caption[]{Top panel: Galactic latitude distribution
of the HII regions of the Synthetic catalog.
The mean Galactic latitude is is $\it{b}$=0.05$^\circ$. Bottom
panel: Galactic longitude distribution.}
\centerline{\includegraphics[width=7.cm,height=5.5cm]{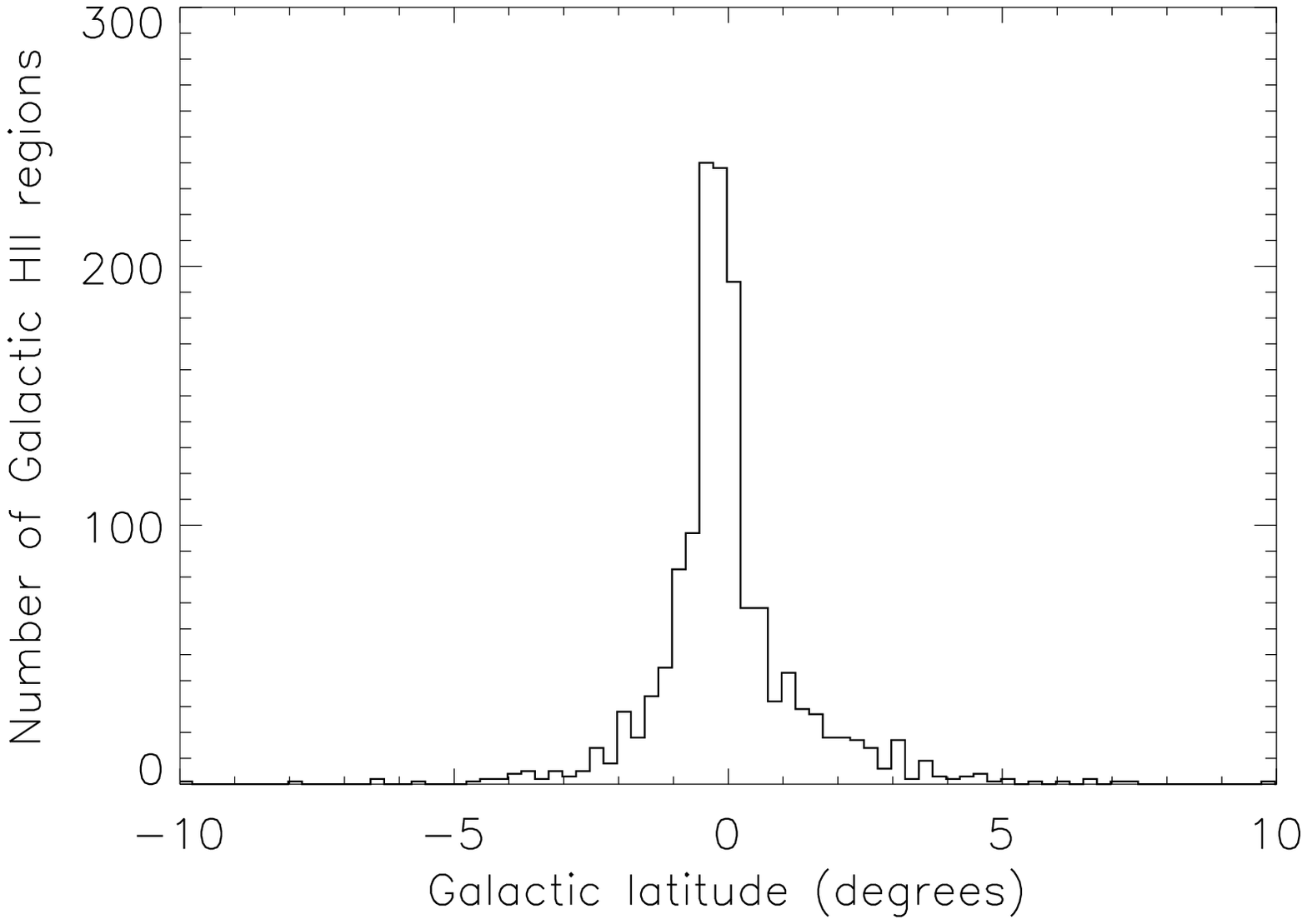}}
\centerline{\includegraphics[width=7cm,height=5.5cm]{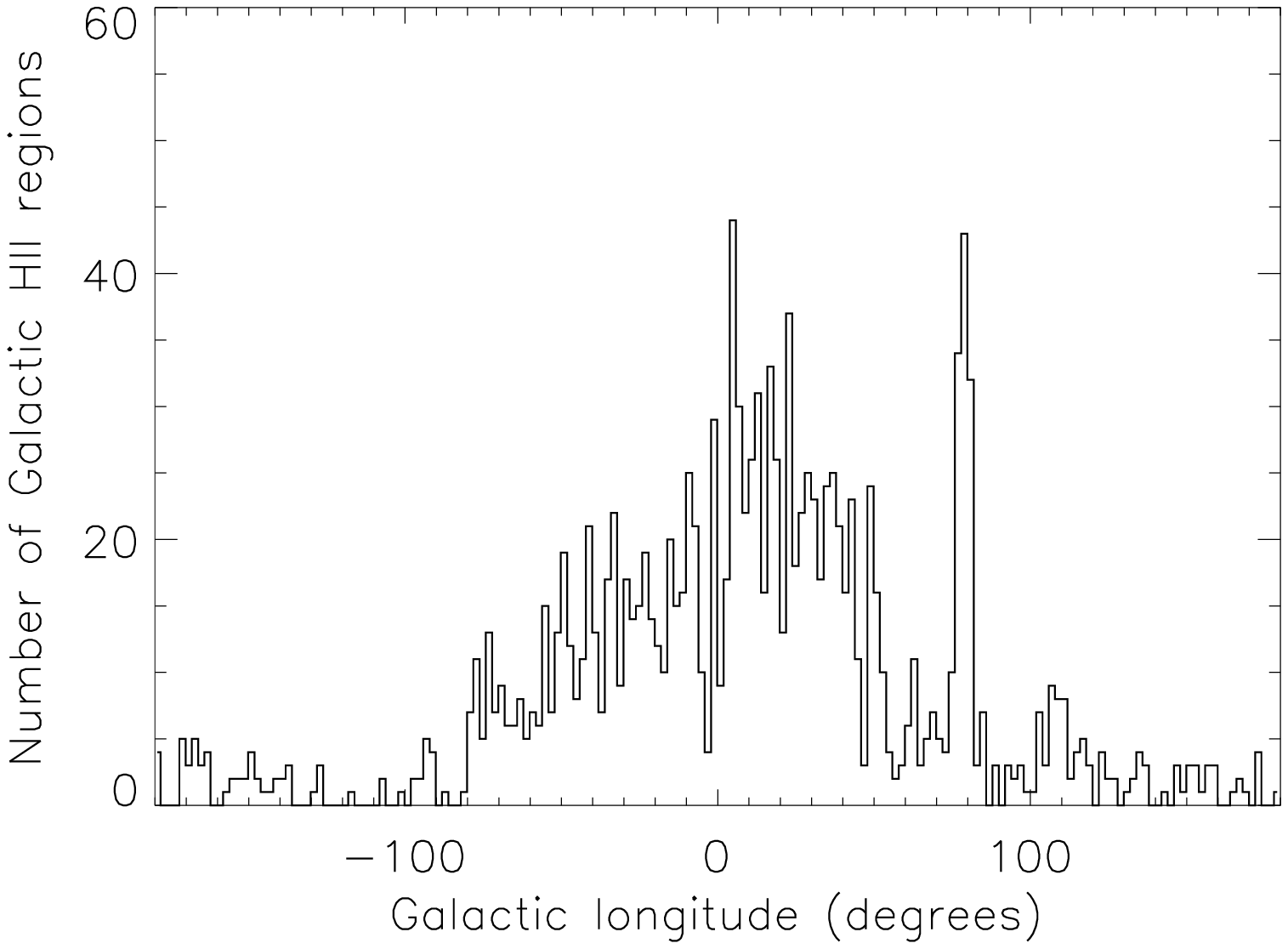}}
\end{figure}

\noindent
The absence of sources at lower flux densities is largely the consequence of
source confusion along the Galactic plane. 
For example, in the region $|\,${\it{l}}$\,| \, < \, 60^\circ$
there are 1200 Synthetic Catalog sources within the Galactic plane 
($|\,${\it{b}}$\,| \, < \, 2^\circ$);
in a 10$^{\prime}$ beam (or a 10$^{\prime}$ diameter source) there is one HII region per
15 beam area which represents a significant confusion level.\\
It is clearly difficult to resolve into individual sources the structure
seen in the central regions of the Galaxy. The majority, $\sim 150$, of the weaker ($<$ 1 Jy) 
sources catalogued by Kuchar $\&$ Clark are in the less complex regions of
the Galaxy such as the anticentre region (quadrants 2 and 3) -- 73 sources -- and 
Galactic latitudes greater than
1$^\circ$  ($|\,${\it{b}}$\,| \, > \, 1^\circ$) in quadrants 1 and 4 -- 19 sources --.

\subsection{Spatial distribution}

\noindent
The Galactic latitude and longitude distribution of the HII regions in the Synthetic
Catalog are shown in Fig~6. A striking feature is the narrow 
distribution in Galactic latitude where the full width at half power is 
0.95$^\circ$ when averaged over all longitudes; the mean Galactic latitude
is b=0.05$^\circ$. This narrow distribution will reflect the distribution
of the O and B stars responsible for the ionization.\\
The main concentration in longitude is at $|$ $\it{l}$ $|$ $<$ 60$^\circ$ where the line
of sight cuts the spiral arms internal to the local arm in which the Sun lies. 
There are also discernable peaks at $\it{l}$ $\sim$ $\pm$ 80$^\circ$ associated with the local
arm; the peak at $\it{l}$ $\sim$ 80$^\circ$ is the Cygnus X region where Wendker 1970
identified 77 HII regions in his survey. There is a clear deficit of sources
in the anticentre region of the Galaxy ($\it{l}$=90$^\circ$--270$^\circ$). Although there is
a less complete coverage in this sector of the Galactic plane as indicated
in Table~1, the large deficit of sources is real. The optical study of
the distribution of HII regions in the outer Galaxy by Fich $\&$ 
Blitz 1984 indicate that they are limited to R $<$ 20 Kpc.

\section{Applications of the Synthetic Catalog}

An extensive data base of radio observations of HII regions has
been distilled into the Synthetic Catalog of 1442 objects at a frequency of 2.7 GHz. 
We consider some uses of this catalog and the associated Master Catalog for
studies of individual HII regions over a range of frequencies and for
CMB studies.

\subsection{Detectability of HII regions}

\noindent
The sensitivity of a particular instrument in Kelvin to an 
HII region of a given flux density expressed in Jansky depends upon
the observing frequency $\nu$, the beamwidth (FWHM),
the sensitivity per second of integration and the observing time $t$
(the rms noise decreases with observing time
$t$ as $t^{-\frac{1}{2}}$). 
The rms noise 
in Jansky per second of integration, $rms_{1s,f}$, is related to the
rms antenna temperature in Kelvin per second of integration,  $rms_{1s,a}$, by:

\begin{eqnarray}
\lefteqn{(rms_{1s,f}/{\rm Jy}) = 2.95 \times 10^{-3} ({\rm FWHM/arcmin})^{2} \cdot} \\
                               &{} &{} (\nu/{\rm GHz})^2 (rms_{1s,a}/{\rm K}) \nonumber
\end{eqnarray}

\noindent
and the same relation holds between the source signal expressed
in terms of flux density or of antenna temperature.
Here the source is assumed to be point-like and observed at the centre
of a Gaussian, symmetric beam. 
Eq.~(1) can be then used to determine the S/N for a point source
of a given flux density. In cases when a source with
a diameter comparable or bigger than the FWHM of the ob\-se\-rving beam is considered, a 
correcting factor to Eq.~(1) is required to compute the S/N. In particular, 
the S/N is $\sim$ 40 to 90 $\%$ the value computed with Eq.~(1) when a source with a
diameter $\sim 2-3$ times the beamwidth or $\sim 1/2-1/3$ the beamwidth, respectively, is considered.
The S/N  represents an easy way to compare the signal produced
by Galactic HII regions with the sensitivity of a 
typical CMB anisotropy experiment. Therefore, we apply the above treatment
to the high resolution satellite ex\-pe\-ri\-ment {\sc Planck} 
by ESA~\footnote{http://planck.esa.nl},
scheduled for launch in 2007. 
{\sc Planck} will observe the entire sky
with a sensitivity at the end of the mission of about 10~$\mu$K per 
resolution element. The beamwidths vary between $\sim 33.6'$ and $5'$
from 30 to 857~GHz respectively. 
For numerical estimates, we consider here
the channels at 30 and 100~GHz which have a nominal
FWHM of $\sim$ 33.6' and 10'. 
For comparison, we make the
same calculation for $COBE$-DMR for which, at 31.5 and 90 GHz the rms
temperature was 150 and 100 $\mu$K and the beamwidth was 7$^\circ$
(Boggess et al. 1992).

\begin{table*}[t]
\caption []{\protect Summary of relative (\%)
errors on quoted fluxes and diameters in the corresponding 
sub-catalogs. Errors marked
by $*$ have been estimated by the authors of
the present paper. Errors marked by $**$ correspond to the indicative
diameters estimated by means of the flux-size correlation in Fig~2.
Further details in Sects.~3.1 and 3.2.}
\begin{center}
\begin{tabular}{lcc}
\hline
\hline
{\hskip 1.1truecm {\it Reference }} & $\%$ \hskip 0.1truecm flux \hskip 0.1truecm error & $\%$
\hskip 0.1truecm size  \hskip 0.1truecm error\\
\hline
Altenhoff et al. 1970$^{a}$  & 10/30 &  6.5--0.4$\times$$\Theta$ ($*$)\\
Altenhoff et al. 1970$^{b}$  & 10/30 & 33.9+2.5$\times$$\Theta$ ($*$) \\
Altenhoff et al. 1979  & 5 & 10   \\
Beard 1966  & 33.5--0.1$\times$S  ($*$) & 300 ($**$)\\  
Beard $\&$ Kerr 1969  & 33.5--0.1$\times$S ($*$) & 37--0.3$\times$$\Theta$ ($*$)\\
Beard et al. 1969  & 33.5--0.1$\times$S ($*$) & 37--0.3$\times$$\Theta$ ($*$) \\
Berlin et al. 1985  & 10 & 72.5 ($*$)\\
Caswell $\&$ Haynes 1987  & 10 & 22.1+2.8$\times$$\Theta$ ($*$)\\
Day et al. 1969  & 33.5--0.1$\times$S ($*$) &  300 ($**$)\\
Day et al. 1970  & 33.5--0.1$\times$S ($*$) &  300 ($**$)\\
Downes et al. 1980  & 5 & 10\\
Felli $\&$ Churchwell 1972  & 15/35 & 77.4 ($*$)\\
F$\ddot{u}$rst et al. 1987  & 10 & 300 ($**$)\\
Goss $\&$ Day 1970  & 33.5--0.1$\times$S ($*$) & 37--0.3$\times$$\Theta$ ($*$) \\
Kuchar $\&$ Clark 1997  & 10/20 & 16/25\\
Mezger $\&$ Henderson 1967  & 23.3($*$) & 20/50\\
Reich et al. 1986 & 10 & 300 ($**$)\\
Reifenstein et al. 1970  & 15 & 3\\
Thomas $\&$ Day 1969a  & 33.5--0.1$\times$S ($*$)  & 300 ($**$)\\
Thomas $\&$ Day 1969b  & 33.5--0.1$\times$S ($*$)  & 300 ($**$)\\
Wendker 1970  & 6 & --73.6+17.6$\times$$\Theta$--0.5$\times$$\Theta^2$\\
Wilson et al. 1970  & 15 & 5\\
Wink et al. 1982  & 5/15 & 73.5--52.1$\times$$\Theta$+11.1$\times$$\Theta^2$ ($*$)\\
Wink et al. 1983  & 30.3--1.9$\times$S+0.1$\times$ S$^2$ ($*$) &
85.8--38.1$\times$$\Theta$+5.2$\times$$\Theta^2$ ($*$)\\
\hline
\hline
\end{tabular}
\end{center}
\end{table*}

\noindent 
The S/N for all the HII regions in the Synthetic Catalog have been
calculated as discussed above; the flux densities were estimated
at each frequency from 2.7 GHz values assuming a flux
density proportional to $\nu^{-0.1}$.
Fig.~7 shows the number of sources per bin of S/N (the distribution
function) and number of sources greater than a certain S/N (cumulative
distribution function) for both {\sc Planck} and $COBE$-DMR at $\sim$ 30 and 
$\sim$ 100 GHz. 

\begin{figure*}[htb]
\mbox{}
\caption[]{Distribution functions (bottom curves) and cumulative distribution functions
(upper curves) at 30 and 100 GHz for {\sc Planck} (dotted lines) and $COBE$-DMR
(dash-dot lines). The $COBE$-DMR angular resolution is 
7$^\circ$ at both frequencies.}
\centerline{\includegraphics[width=5.5cm,height=7.8cm,angle=90]{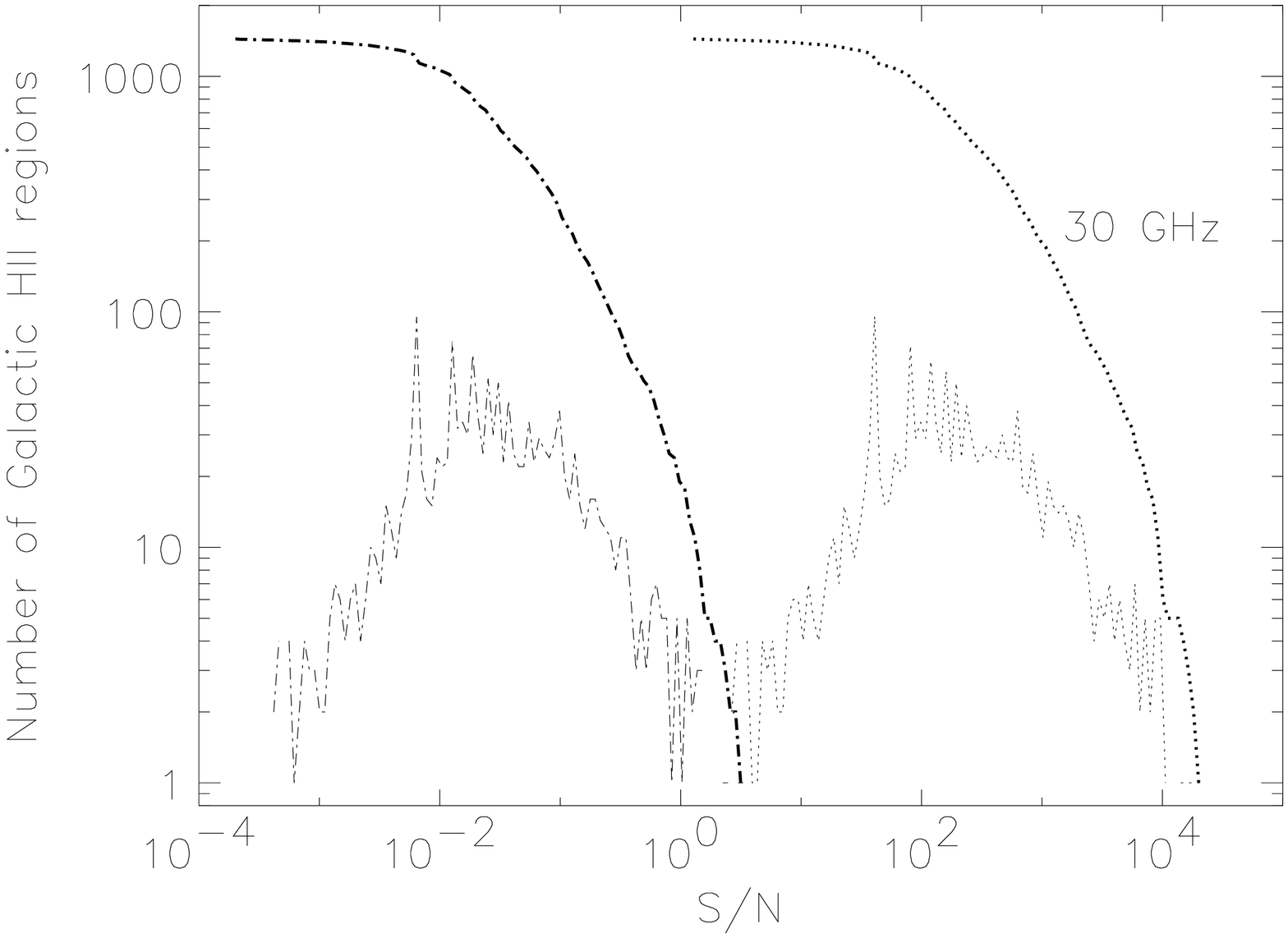}
            \includegraphics[width=5.5cm,height=7.8cm,angle=90]{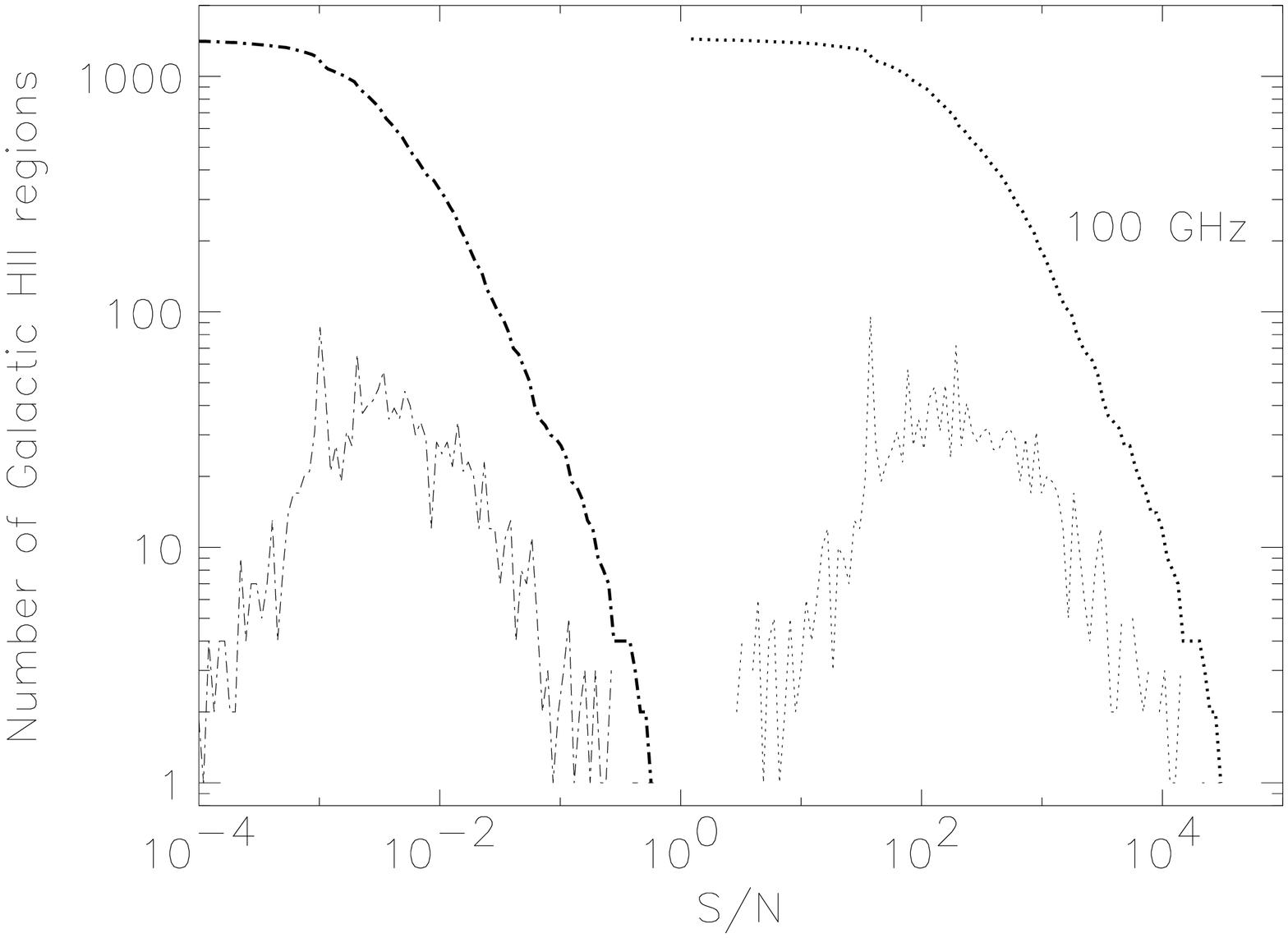}}
\end{figure*}

\noindent
From Fig.~7 it is clear that 
the vast majority of sources in the Synthetic Catalog
produces a signal which highly exceeds the detection threshold
of the instrument. Therefore, not only all the Synthetic Catalog
HII regions will be detected but also $80\%$ of these sources will have
a S/N $ >$ 100 (while the weakest sources, with S $\sim$ 10 mJy, will
be seen with a S/N $\sim$ 3). 
As a consequence, {\sc Planck}
high sensitivity and high resolution should allow to
significantly entend the existing HII regions data base.
It is important to emphasize that
none of the individual HII regions would have been seen by $COBE$-DMR whose
flux sensitivity according to Eq.~(1) is $6 \times 10^{3}$ less than {\sc Planck}
at 30 GHz and $4 \times 10^{4}$ less at 100 GHz.

\begin{figure*}[htb]
\mbox{}
\caption[]{Simulated full sky map of the antenna temperature signal at 30~GHz
produced by the Synthetic Catalog HII regions for an
instrument with a FWHM of 33.6'. The map is displayed in all-sky
mollweide projection.
The color bar shows the minimun and maximum of the
pixels in the map with a positive signal. The
signal looks particularly bright with peaks of tens of mK.
See also the text.}
\centerline{\includegraphics[width=7cm, height=13.5cm, angle=-90]{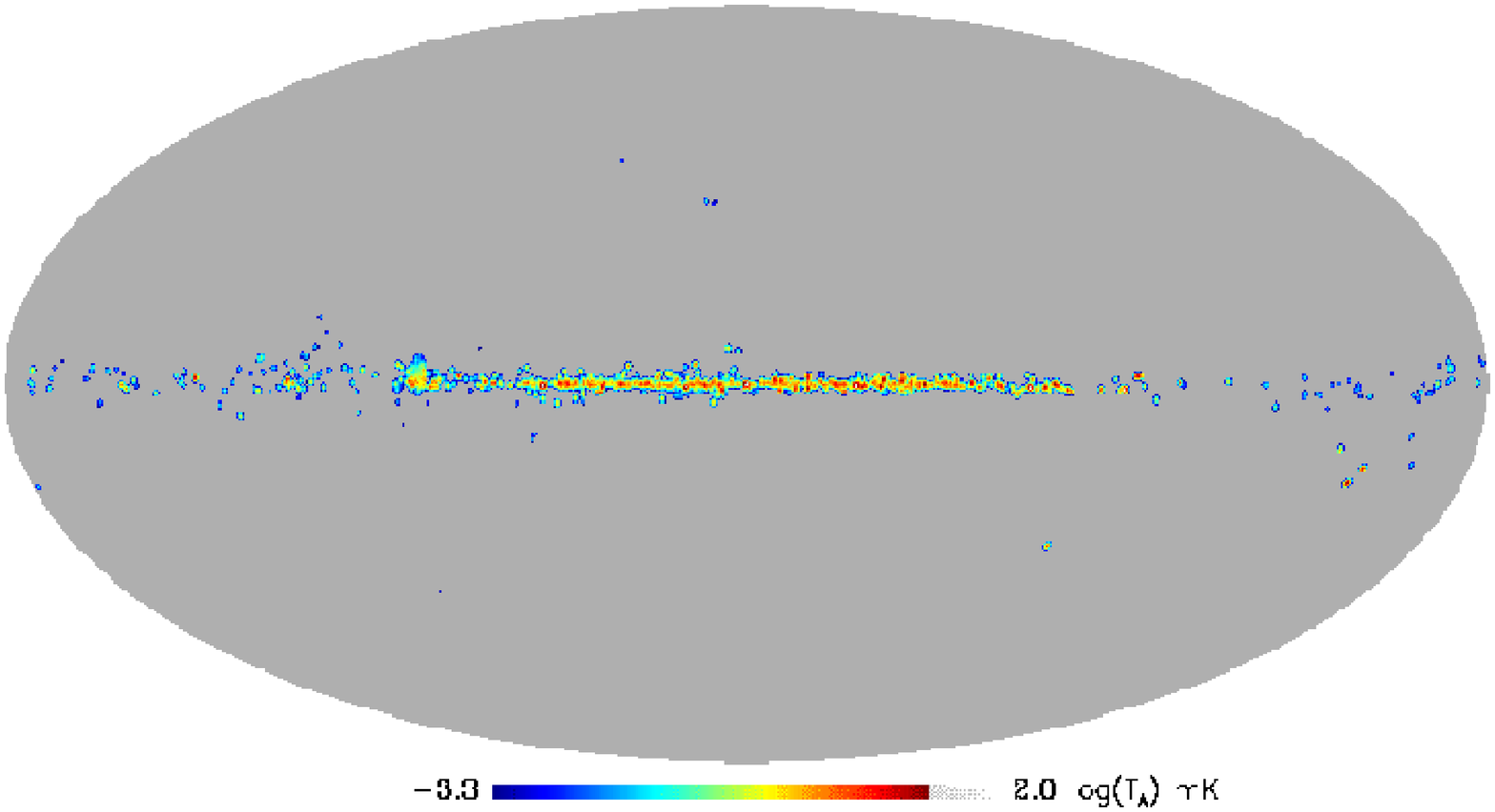}}
\end{figure*}

\subsection{Radio/millimeter studies of HII regions}

\noindent
The ionizing radiation from the central O/B stars produces the HII region which
emits through the free-free mechanism from radio to submillimetric wavelenghts; 
the surrounding dust is also heated by radiation originally derived from the
central stars and as a consequence radiates at submillimeter and IR 
wavelenghts. The flux density of the radiation from these 
two components was found to be equal in the
wavelenght range 1-3 mm (100 to 300 GHz) by Kurtz et al. 1994 for a selection of compact
HII regions. However, many interesting questions remain to be
resolved in the physical relationship between the HII region
and the radiation-heated dust cloud in which it is embedded. The Synthetic
Catalog provides a rich source of HII regions for further
study. Fig.~7 shows that many hundreds of HII regions will be detectable 
at high sen\-si\-ti\-vi\-ty and with adequate resolution over the frequency range 
30 to 857 GHz by {\sc Planck}. Thus, this may represent a good chance
for a comparative study of the far-IR and radio continuum
morphology of Galactic HII regions.\\
Moreover, source identification in IR experiments 
such as DENIS (Epchetein et al., 1994) and 2MASS (Kleinmann, 1992) 
can benefit from the crosscheck with the Synthetic Catalog. In fact, 
for these kind of experiments a major problem is the association 
of an observed source with a bright IR Galactic object like an
HII region or an ultracompact HII region rather than with an
external Galaxy.

\subsection{Use of the Synthetic Catalog in CMB studies}

\noindent
We will consider in this subsection several applications
of the Synthetic Catalog in CMB studies. The first use we discuss 
is the
production of maps of the integrated free-free emission as seen with the
angular resolution of an instrument such as {\sc Planck} at each
observing frequency. Free-free emission dominates the Galactic plane
signal at least over the range 30 to 100 GHz. Fig.~8 shows the Galactic
plane emission resulting from HII regions as seen by {\sc Planck} at
30 GHz where the beamwidth is 33.6'.\\
To make the map in Fig.~8, we have implemented a code which simulates 
the contribution of each source in the Synthetic Catalog at a given angular resolution 
by assuming a symmetric Gaussian profile for the source
brightness distribution 
and numerically convolves, in real space,
the relevant part of the map -- obtained after considering
the contribution from all the sources in the catalog --
with another symmetric Gaussian profile having the instrument FWHM.
The final map is generated in HEALPix
(Hierarchical Equal Area and IsoLatitude Pixelization
of the Sphere, by G\`orski et al. 1998).  Fig.~9 shows the strong Galactic
centre and Cygnus X regions. Antenna temperature as high
as $\sim$ 50 mK are seen.\\
Moreover, HII regions, as delineated in Fig.~8, can in principle represent 
a significant contribution -- comparable to that produced by synchrotron and
dust emission -- to one of the most critical
spurious effects in CMB surveys,
the so-called straylight contamination through their integrated signal in the sidelobes 
(Burigana et al. 2001) which needs to be properly evaluated.
The  optical design of {\sc Planck} and similar
mapping instruments must be optimized
to minimize this effect. In any case, the residual straylight 
should be taken into account in the data analysis.\\
HII regions also have a part to play in CMB imaging experiments 
as suitable calibrators and as probes of the pointing and
beamshape. HII regions are non-variable and have quite a well-known
spectrum which makes them va\-lua\-ble calibrators along with planets 
and the CMB dipole (Bersanelli et al. 1997). In addition, for space missions, they provide
auxiliary data for inflight beam re\-con\-stru\-ction and
pointing by complementing the information from planet transits
(Burigana et al. 2002) and from the interplay between amplitudes
and phases of CMB signal with the instrumental noise (Chiang et al. 2002).
In particular, the Chiang et al. method, although not
requiring the use of bright sources, allows the
reconstruction of the beam ellipticity only
in the beam central regions while the Burigana et al. technique allows the complete
reconstruction of the beam shape down to a level of -25 dB but makes
use of non-variable, bright, compact sources. However, despite being originally 
conceived for planets, the Burigana et al. method 
can be easily extended to other classes of sources which also
enable to increase the number of transits over the space mission lifetime.
The Synthetic Catalog contains 36 HII regions which have a flux density
 $\ge$ 30 Jy at 30 GHz and a diameter $\le$ than 5'. Accurate flux densities
ad positions can be determined from ground-based aperture synthetis
observations.\\
Finally, we point out that the angular extension of 
a typical Galactic HII region represents an intermediate
case between point sources and diffuse foregrounds for which  
component separation tools have been specifically designed 
(see, e.g., Maino et al. 2001, and references therein). The fluctuations produced by such
extended sources and the capability of existing component
separation methods to handle with them will have to be furtherly 
investigated in the next years.

\noindent
\begin{figure*}[htb]
\centerline{\includegraphics[width=7.5cm,height=6.9cm]{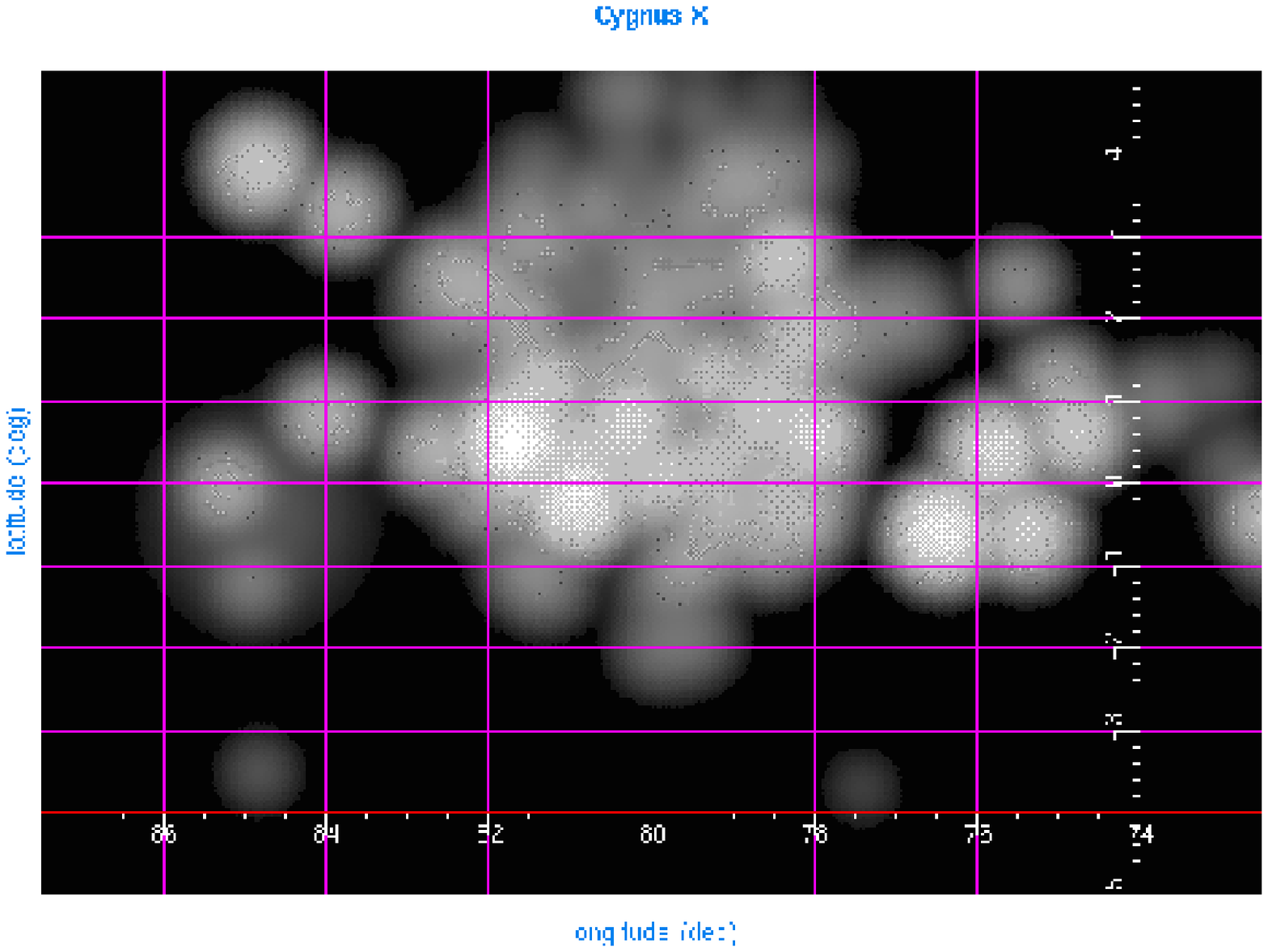}  
           \includegraphics[width=7.5cm,height=7cm]{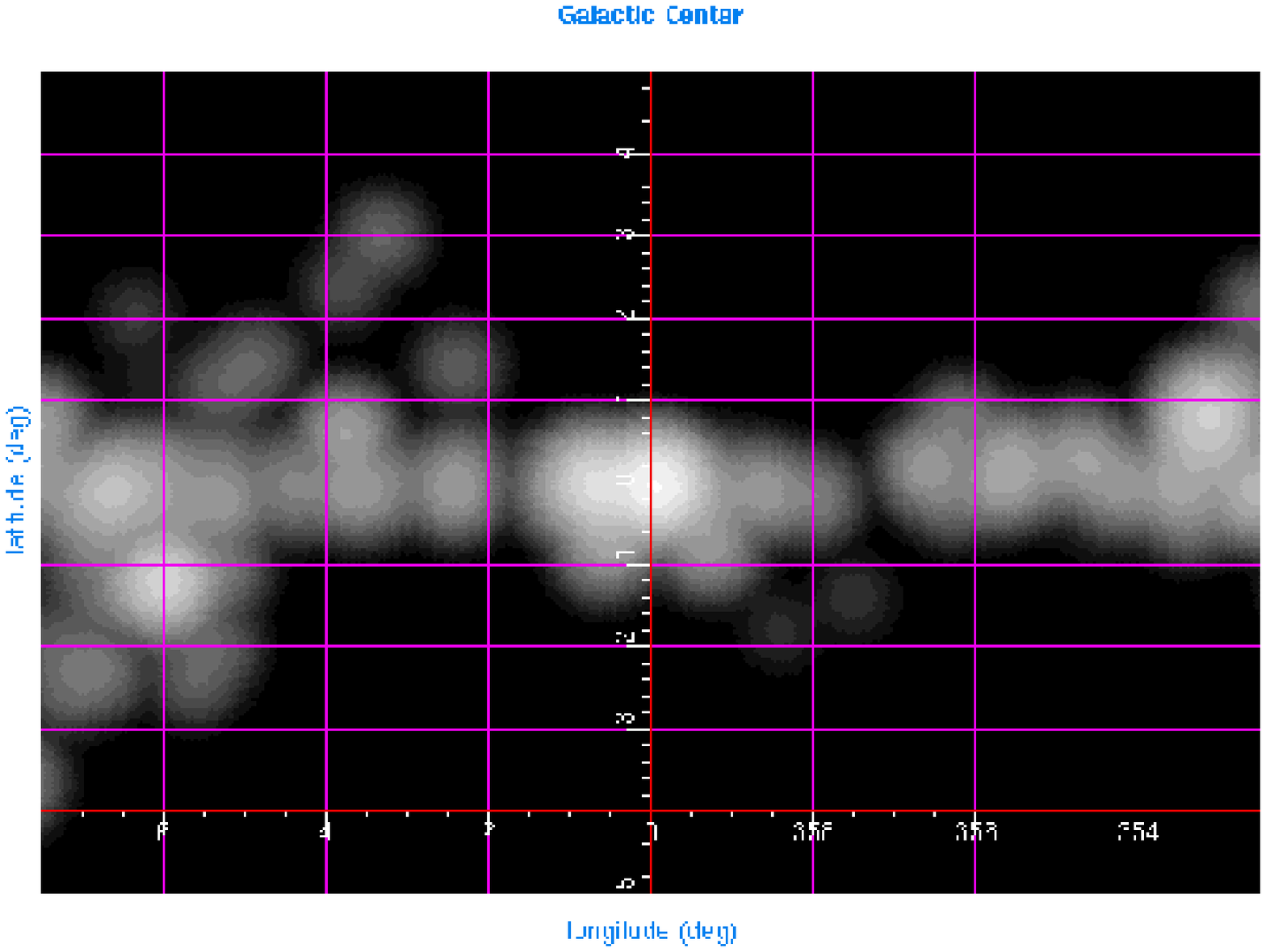}}
\hspace*{2.truecm}{\includegraphics[width=6.5cm,height=1cm]{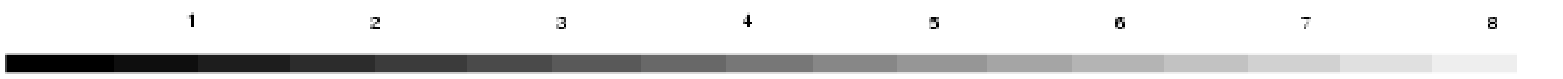}}
\hspace*{1.truecm}{\includegraphics[width=6.5cm,height=1cm,angle=0]{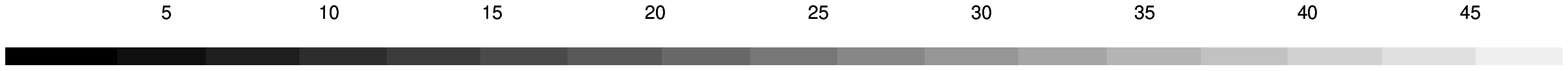}}
\caption[]{Simulated maps of the Cygnus X region 
(left panel) and of the Galactic Center region (right panel) 
at 30~GHz with a FWHM resolution of $33.6'$. The units of
the color bar are mK. Minimum and maximum refer to 
positive pixel signal values.} 
\end{figure*}

\section{Conclusions}
We have collected radio data of Galactic HII regions 
from 24 published works and built a self-consistent
database of 1442 sources. This work has resulted in the
construction of what we have called Master Catalog which
consists in 11 different sub-catalogs storing the
original information from the source references. In particular,
the sub-catalogs list flux densities and diameters as well
as radio line ve\-lo\-ci\-ties, line widths and line
temperatures; errors are given on these
quantities. \\
From this large data set observed at a range of frequencies and
beamwidths, we have produced a readibily accessible Synthetic Catalog 
giving the flux density, diameter and velocity (where available)
at 2.7 GHz, the frequency where most data are given. Errors on these
observed parameters are derived using the procedures discussed in
Sect.~3. \\
It should be emphasized that the Catalog is a compilation of published radio
data on HII regions; it is not a complete survey down to the
faintest flux density level listed. We argue in Sect.~4 that it is probably
complete to a flux density of 1 Jy. Nevertheless, the Catalog
provides an up to date finding list for the brightest HII regions 
available for further study at microwave and submillimeter wavelenghts. The
Synthetic Catalog is also particularly relevant to future
high sensitivity CMB mapping projects such as {\sc Planck}. We have
shown in Sect.~5 that the HII region Galactic distribution can contribute
to the straylight radiation, while individual compact bright HII regions
have a role as calibrators and as pointing and beamshape indicators.\\
The Master Catalog and the Synthetic Catalog are a\-va\-ila\-ble via
ftp at: cdsarc.u-strasbg.fr.\\
Color figures can be obtained via e-mail request to 
paladini@sissa.it.

\begin{acknowledgements}
We wish to thank C.~Witebsky for providing us with
the $COBE$-HII compilation and T.A.~Kuchar and F.O.~Clark 
for making their radio and IRAS data available prior to publication. 
We are also grateful to the CATS group for 
a constructive support to this work and to 
G.~De~Zotti and H.J.~Wendker for useful
discussions and suggestions. We thank L. Cambrecy for valuable 
comments to our work.
R.~Paladini acknowledges financial support from a N.A.T.O. International
Exchange Scientific Programme grant and from a Marie Curie
Training Site fellowship.
We gratefully acknowledge K.M.~G\'orski and all the people
involved in the realization of the tools of HEALPix pixelisation.
\end{acknowledgements}

\centerline{\bf Appendix~I: the Master Catalog}

\vspace*{0.2truecm}
\noindent
The Master Catalog contains the original data taken
from the source references of Table~1. Due
to the extensive amount of information, we have
divided for convenience the Master Catalog into
11 sub-catalogs. We give hereafter the description
of the structure of each of these compilations while,
for a general overview about the content of the sub-catalogs, we
refer the reader to Sec.~2.5.\\
As mentioned in the main text (Sect.~2.5), each sub-catalog
(except for Sub-catalog~1)
has 37 columns and 1442 entries corresponding to the
total number of sources. The columns are in order of
increasing frequency of observation and, at each
frequency, the columns are in alphabetical order of the references
in Table~1. In all sub-catalogs, a null entry corresponds to
no information available from the source reference.

\vspace*{0.2truecm}
\noindent
Sub-catalog 1 lists only source coordinates, notes about the
environment and radio/optical couterparts. For its content
specificy, it presents a peculiar structure 
with respect to the other sub-catalogs. In details:\\
- Col. 1: source-numbering (records from 1 to 1442)\\
- Col. 2-3: Galactic coordinates, l and b\\
- Col. 4-6: celestial coordinates: RA - J2000\\
- Col. 7-9: celestial coordinate: DEC - J2000\\  
- Col. 10: general remarks - C= complex field, S= strong
source nearby, X=strong source nearby ($>$ 10 Jy), radio or
optical counterpart. This flags follow the definitions given 
by Kuchar $\&$
Clark 1997 according to which a source is in a complex field 
when there are two or more
sources within either $2\times\Theta_{obs}$ (i.e, four source
radii) or $2\times\Theta_{beam}$, whatever is
larger {\footnote{In Kuchar $\&$ Clark's paper, fluxes and
angular sizes are measured directly
from the survey images.}}, or with a strong and/or much
stronger source nearby.\\
As for the counterpart in other wavebands, for the case of
the radio identifications: Ke refers to Kesteven 1968; NRAO
to Pauliny-Toth 1966; W to Westerhout 1958; 3C to Third
Cambridge Catalog (Bennett 1962) and 4C to Fourth Cambridge
Catalog (Gower et al. 1967 - Pilkington et al. 1965).\\
For the optical identifications, the M\'arsalkov\'a Catalog
1974 has been
consulted. In this case: BBW refers to Bok-Bester-Wade
1955; DWB to Dickel-Wendker-Bieritz
1969; Ge (a) to Georgelin-Georgelin 1970a; Ge (b) to
Georgelin-Georgelin 1970b;
Ge (c) to Georgelin-Georgelin 1970c; G to Gum 1955; H to
Hoffleit 1953;
RCW to Rodgers-Campbell-Whiteoak 1960; S to Sharpless
1959. Moreover, M stands
for Messier Catalog; NGC for Dreyers'New General Catalog and
IC for the Index Catalog.

\vspace*{0.1truecm}
\noindent
Sub-catalog 2 to sub-catalog 11 have the following
structure:\\
- Col. 1: source-numbering (records from 1 to 1442)\\
- Col. 2-3: Galactic coordinates, l and b\\
- Col. 4-6: celestial coordinates: RA - J2000\\ 
- Col. 7-9: celestial coordinate: DEC - J2000\\
- Col. 10-11: 1.4 GHz - Altenhoff et al. (1970),
  Felli $\&$ Churchwell (1972)\\
- Col. 12-23: 2.7 GHz - Altenhoff et al. 1970,
  Beard 1966, Beard $\&$ Kerr
  1969, Beard et al. 1969, Day et al. 1969, Day et al.
  1970, F$\ddot{u}$rst et al. 1987,  
  Goss $\&$ Day 1970, Reich et al. 1986, Thomas $\&$ Day
  1969a, Thomas $\&$ Day 1969b,
  Wendker 1970\\
- Col. 24: 3.9 GHz - Berlin et al. 1985\\
- Col. 25-26: 4.8 GHz - Kuchar $\&$ Clark 1997\\
- Col. 27-34: 5 GHz - Altenhoff et al. 1970,
  Altenhoff et al. 1979, Caswell
  $\&$ Haynes 1987, Downes et al. 1980, Mezger $\&$ Henderson 1967, Reifenstein et al. 1970, 
  Wilson et al. 1970, Wink et al. 1982\\
- Col. 35: 14.7 GHz - Wink et al. 1983\\
- Col. 36: 15 GHz - Wink et al. 1982\\
- Col. 37: 86 GHz - Wink et al. 1982.\\

\noindent
Note on sub-catalog 2 and sub-catalog 4: these sub-catalogs,
including data on flux and angular diameters, always list 
values corrected for the instrument beam. Their complementary sub-catalogs
are sub-catalog 3 and 5 quoting relative (\%) errors.\\

\vspace*{0.2truecm}
\noindent
Note on sub-catalog 6 to 11: these sub-catalogs list line velocity data. 
Observations may refer
to frequencies other than those for the continuum data. In 
particular:\\
- Col. 25-26: Kuchar $\&$ Clark quote continuum data at 4.85 GHz and  
  line data from Reifenstein et al. 1970
  and from Wilson et al. 1970 at 5 GHz (H109$\alpha$) 
  and from Lockman 1989 at 10 GHz (H85$\alpha$, H87$\alpha$ and H88$\alpha$)\\
- Col. 28: Altenhoff et al. 1979
  quotes continuum data at 5 GHz while line data
  are taken from Lockman 1989 at 10 GHz (H85$\alpha$, H87$\alpha$ and H88$\alpha$)\\
- Col. 29-33: Caswell $\&$ Haynes 1987, Downes et al. 1980, Mezger $\&$
  Henderson 1967, Reifenstein
  et al. 1970, Wilson et al. 1970 quote both
  continuum and line data at 5 GHz (H109$\alpha$, H110$\alpha$)\\
- Col. 34-36-37: Wink et al. 1982 quotes continuum data 
  at 5/15/86 GHz and line data at 8.9/14.7 GHz 
  (H90$\alpha$ and H76$\alpha$).

\end{document}